\documentstyle[editedvolume]{crckapb} 
\def\PsfigVersion{1.10}
\def\setDriver{\DvipsDriver} 
\ifx\undefined\psfig\else\endinput\fi
\let\LaTeXAtSign=\@
\let\@=\relax
\edef\psfigRestoreAt{\catcode`\@=\number\catcode`@\relax}
\catcode`\@=11\relax
\newwrite\@unused
\def\ps@typeout#1{{\let\protect\string\immediate\write\@unused{#1}}}

\def\DvipsDriver{
	\ps@typeout{psfig/tex \PsfigVersion -dvips}
\def\PsfigSpecials{\DvipsSpecials} 	\def\ps@dir{/}
\def\ps@predir{} }
\def\OzTeXDriver{
	\ps@typeout{psfig/tex \PsfigVersion -oztex}
	\def\PsfigSpecials{\OzTeXSpecials}
	\def\ps@dir{:}
	\def\ps@predir{:}
	\catcode`\^^J=5
}

\def\figurepath{./:}

\def\DoPaths#1{\expandafter\EachPath#1\stoplist}
\def\leer{}
\def\EachPath#1:#2\stoplist{
  \ExistsFile{#1}{\SearchedFile}
  \ifx#2\leer
  \else
    \expandafter\EachPath#2\stoplist
  \fi}
\def\ps@dir{/}
\def\ExistsFile#1#2{%
   \openin1=\ps@predir#1\ps@dir#2
   \ifeof1
       \closein1
   \else
       \closein1
        \ifx\ps@founddir\leer
           \edef\ps@founddir{#1}
        \fi
   \fi}
\def\get@dir#1{%
  \def\ps@founddir{}
  \def\SearchedFile{#1}
  \DoPaths\figurepath
}

\def\@nnil{\@nil}
\def\@empty{}
\def\@psdonoop#1\@@#2#3{}
\def\@psdo#1:=#2\do#3{\edef\@psdotmp{#2}\ifx\@psdotmp\@empty \else
    \expandafter\@psdoloop#2,\@nil,\@nil\@@#1{#3}\fi}
\def\@psdoloop#1,#2,#3\@@#4#5{\def#4{#1}\ifx #4\@nnil \else
       #5\def#4{#2}\ifx #4\@nnil \else#5\@ipsdoloop #3\@@#4{#5}\fi\fi}
\def\@ipsdoloop#1,#2\@@#3#4{\def#3{#1}\ifx #3\@nnil 
       \let\@nextwhile=\@psdonoop \else
      #4\relax\let\@nextwhile=\@ipsdoloop\fi\@nextwhile#2\@@#3{#4}}
\def\@tpsdo#1:=#2\do#3{\xdef\@psdotmp{#2}\ifx\@psdotmp\@empty \else
    \@tpsdoloop#2\@nil\@nil\@@#1{#3}\fi}
\def\@tpsdoloop#1#2\@@#3#4{\def#3{#1}\ifx #3\@nnil 
       \let\@nextwhile=\@psdonoop \else
      #4\relax\let\@nextwhile=\@tpsdoloop\fi\@nextwhile#2\@@#3{#4}}
\ifx\undefined\fbox
\newdimen\fboxrule
\newdimen\fboxsep
\newdimen\ps@tempdima
\newbox\ps@tempboxa
\long\def\fbox#1{\leavevmode\setbox\ps@tempboxa\hbox{#1}\ps@tempdima\fboxrule
    \advance\ps@tempdima \fboxsep \advance\ps@tempdima \dp\ps@tempboxa
   \hbox{\lower \ps@tempdima\hbox
  {\vbox{\hrule height \fboxrule
          \hbox{\vrule width \fboxrule \hskip\fboxsep
          \vbox{\vskip\fboxsep \box\ps@tempboxa\vskip\fboxsep}\hskip 
                 \fboxsep\vrule width \fboxrule}
                 \hrule height \fboxrule}}}}
\fi
\newread\ps@stream
\newif\ifnot@eof       
\newif\if@noisy        
\newif\if@atend        
\newif\if@psfile       
{\catcode`\%=12\global\gdef\epsf@start{
\def\epsf@PS{PS}
\def\epsf@getbb#1{%
\openin\ps@stream=\ps@predir#1
\ifeof\ps@stream\ps@typeout{Error, File #1 not found}\else
   {\not@eoftrue \chardef\other=12
    \def\do##1{\catcode`##1=\other}\dospecials \catcode`\ =10
    \loop
       \if@psfile
	  \read\ps@stream to \epsf@fileline
       \else{
	  \obeyspaces
          \read\ps@stream to \epsf@tmp\global\let\epsf@fileline\epsf@tmp}
       \fi
       \ifeof\ps@stream\not@eoffalse\else
       \if@psfile\else
       \expandafter\epsf@test\epsf@fileline:. \\%
       \fi
          \expandafter\epsf@aux\epsf@fileline:. \\%
       \fi
   \ifnot@eof\repeat
   }\closein\ps@stream\fi}%
\long\def\epsf@test#1#2#3:#4\\{\def\epsf@testit{#1#2}
			\ifx\epsf@testit\epsf@start\else
\ps@typeout{Warning! File does not start with `\epsf@start'.  It may not be a PostScript file.}
			\fi
			\@psfiletrue} 
{\catcode`\%=12\global\let\epsf@percent=
\long\def\epsf@aux#1#2:#3\\{\ifx#1\epsf@percent
   \def\epsf@testit{#2}\ifx\epsf@testit\epsf@bblit
	\@atendfalse
        \epsf@atend #3 . \\%
	\if@atend	
	   \if@verbose{
		\ps@typeout{psfig: found `(atend)'; continuing search}
	   }\fi
        \else
        \epsf@grab #3 . . . \\%
        \not@eoffalse
        \global\no@bbfalse
        \fi
   \fi\fi}%
\def\epsf@grab #1 #2 #3 #4 #5\\{%
   \global\def\epsf@llx{#1}\ifx\epsf@llx\empty
      \epsf@grab #2 #3 #4 #5 .\\\else
   \global\def\epsf@lly{#2}%
   \global\def\epsf@urx{#3}\global\def\epsf@ury{#4}\fi}%
\def\epsf@atendlit{(atend)} 
\def\epsf@atend #1 #2 #3\\{%
   \def\epsf@tmp{#1}\ifx\epsf@tmp\empty
      \epsf@atend #2 #3 .\\\else
   \ifx\epsf@tmp\epsf@atendlit\@atendtrue\fi\fi}

\chardef\psletter = 11 
\chardef\other = 12

\newif \ifdebug 
\newif\ifc@mpute 
\c@mputetrue 

\let\then = \relax
\def\r@dian{pt }
\let\r@dians = \r@dian
\let\dimensionless@nit = \r@dian
\let\dimensionless@nits = \dimensionless@nit
\def\internal@nit{sp }
\let\internal@nits = \internal@nit
\newif\ifstillc@nverging
\def \Mess@ge #1{\ifdebug \then \message {#1} \fi}

{ 
	\catcode `\@ = \psletter
	\gdef \nodimen {\expandafter \n@dimen \the \dimen}
	\gdef \term #1 #2 #3%
	       {\edef \t@ {\the #1}
		\edef \t@@ {\expandafter \n@dimen \the #2\r@dian}%
		\t@rm {\t@} {\t@@} {#3}%
	       }
	\gdef \t@rm #1 #2 #3%
	       {{%
		\count 0 = 0
		\dimen 0 = 1 \dimensionless@nit
		\dimen 2 = #2\relax
		\Mess@ge {Calculating term #1 of \nodimen 2}%
		\loop
		\ifnum	\count 0 < #1
		\then	\advance \count 0 by 1
			\Mess@ge {Iteration \the \count 0 \space}%
			\Multiply \dimen 0 by {\dimen 2}%
			\Mess@ge {After multiplication, term = \nodimen 0}%
			\Divide \dimen 0 by {\count 0}%
			\Mess@ge {After division, term = \nodimen 0}%
		\repeat
		\Mess@ge {Final value for term #1 of 
				\nodimen 2 \space is \nodimen 0}%
		\xdef \Term {#3 = \nodimen 0 \r@dians}%
		\aftergroup \Term
	       }}
	\catcode `\p = \other
	\catcode `\t = \other
	\gdef \n@dimen #1pt{#1} 
}

\def \Divide #1by #2{\divide #1 by #2} 

\def \Multiply #1by #2
       {{
	\count 0 = #1\relax
	\count 2 = #2\relax
	\count 4 = 65536
	\Mess@ge {Before scaling, count 0 = \the \count 0 \space and
			count 2 = \the \count 2}%
	\ifnum	\count 0 > 32767 
	\then	\divide \count 0 by 4
		\divide \count 4 by 4
	\else	\ifnum	\count 0 < -32767
		\then	\divide \count 0 by 4
			\divide \count 4 by 4
		\else
		\fi
	\fi
	\ifnum	\count 2 > 32767 
	\then	\divide \count 2 by 4
		\divide \count 4 by 4
	\else	\ifnum	\count 2 < -32767
		\then	\divide \count 2 by 4
			\divide \count 4 by 4
		\else
		\fi
	\fi
	\multiply \count 0 by \count 2
	\divide \count 0 by \count 4
	\xdef \product {#1 = \the \count 0 \internal@nits}%
	\aftergroup \product
       }}

\def\r@duce{\ifdim\dimen0 > 90\r@dian \then   
		\multiply\dimen0 by -1
		\advance\dimen0 by 180\r@dian
		\r@duce
	    \else \ifdim\dimen0 < -90\r@dian \then  
		\advance\dimen0 by 360\r@dian
		\r@duce
		\fi
	    \fi}

\def\Sine#1%
       {{%
	\dimen 0 = #1 \r@dian
	\r@duce
	\ifdim\dimen0 = -90\r@dian \then
	   \dimen4 = -1\r@dian
	   \c@mputefalse
	\fi
	\ifdim\dimen0 = 90\r@dian \then
	   \dimen4 = 1\r@dian
	   \c@mputefalse
	\fi
	\ifdim\dimen0 = 0\r@dian \then
	   \dimen4 = 0\r@dian
	   \c@mputefalse
	\fi
	\ifc@mpute \then
		\divide\dimen0 by 180
		\dimen0=3.141592654\dimen0
		\dimen 2 = 3.1415926535897963\r@dian 
		\divide\dimen 2 by 2 
		\Mess@ge {Sin: calculating Sin of \nodimen 0}%
		\count 0 = 1 
		\dimen 2 = 1 \r@dian 
		\dimen 4 = 0 \r@dian 
		\loop
			\ifnum	\dimen 2 = 0 
			\then	\stillc@nvergingfalse 
			\else	\stillc@nvergingtrue
			\fi
			\ifstillc@nverging 
			\then	\term {\count 0} {\dimen 0} {\dimen 2}%
				\advance \count 0 by 2
				\count 2 = \count 0
				\divide \count 2 by 2
				\ifodd	\count 2 
				\then	\advance \dimen 4 by \dimen 2
				\else	\advance \dimen 4 by -\dimen 2
				\fi
		\repeat
	\fi		
			\xdef \sine {\nodimen 4}%
       }}

\def\Cosine#1{\ifx\sine\UnDefined\edef\Savesine{\relax}\else
		             \edef\Savesine{\sine}\fi
	{\dimen0=#1\r@dian\advance\dimen0 by 90\r@dian
	 \Sine{\nodimen 0}
	 \xdef\cosine{\sine}
	 \xdef\sine{\Savesine}}}	      

\def\psdraft{
	\def\@psdraft{0}
}
\def\psfull{
	\def\@psdraft{100}
}

\psfull

\newif\if@scalefirst
\def\psscalefirst{\@scalefirsttrue}
\def\psrotatefirst{\@scalefirstfalse}
\psrotatefirst

\newif\if@draftbox
\def\psnodraftbox{
	\@draftboxfalse
}
\def\psdraftbox{
	\@draftboxtrue
}
\@draftboxtrue

\newif\if@prologfile
\newif\if@postlogfile
\def\pssilent{
	\@noisyfalse
}
\def\psnoisy{
	\@noisytrue
}
\psnoisy
\newif\if@bbllx
\newif\if@bblly
\newif\if@bburx
\newif\if@bbury
\newif\if@height
\newif\if@width
\newif\if@rheight
\newif\if@rwidth
\newif\if@angle
\newif\if@clip
\newif\if@verbose
\def\@p@@sclip#1{\@cliptrue}
\newif\if@decmpr
\def\@p@@sfigure#1{\def\@p@sfile{null}\def\@p@sbbfile{null}\@decmprfalse
   \openin1=\ps@predir#1
   \ifeof1
	\closein1
	\get@dir{#1}
	\ifx\ps@founddir\leer
		\openin1=\ps@predir#1.bb
		\ifeof1
			\closein1
			\get@dir{#1.bb}
			\ifx\ps@founddir\leer
				\ps@typeout{Can't find #1 in \figurepath}
			\else
				\@decmprtrue
				\def\@p@sfile{\ps@founddir\ps@dir#1}
				\def\@p@sbbfile{\ps@founddir\ps@dir#1.bb}
			\fi
		\else
			\closein1
			\@decmprtrue
			\def\@p@sfile{#1}
			\def\@p@sbbfile{#1.bb}
		\fi
	\else
		\def\@p@sfile{\ps@founddir\ps@dir#1}
		\def\@p@sbbfile{\ps@founddir\ps@dir#1}
	\fi
   \else
	\closein1
	\def\@p@sfile{#1}
	\def\@p@sbbfile{#1}
   \fi
}
\def\@p@@sfile#1{\@p@@sfigure{#1}}
\def\@p@@sbbllx#1{
		\@bbllxtrue
		\dimen100=#1
		\edef\@p@sbbllx{\number\dimen100}
}
\def\@p@@sbblly#1{
		\@bbllytrue
		\dimen100=#1
		\edef\@p@sbblly{\number\dimen100}
}
\def\@p@@sbburx#1{
		\@bburxtrue
		\dimen100=#1
		\edef\@p@sbburx{\number\dimen100}
}
\def\@p@@sbbury#1{
		\@bburytrue
		\dimen100=#1
		\edef\@p@sbbury{\number\dimen100}
}
\def\@p@@sheight#1{
		\@heighttrue
		\dimen100=#1
   		\edef\@p@sheight{\number\dimen100}
}
\def\@p@@swidth#1{
		\@widthtrue
		\dimen100=#1
		\edef\@p@swidth{\number\dimen100}
}
\def\@p@@srheight#1{
		\@rheighttrue
		\dimen100=#1
		\edef\@p@srheight{\number\dimen100}
}
\def\@p@@srwidth#1{
		\@rwidthtrue
		\dimen100=#1
		\edef\@p@srwidth{\number\dimen100}
}
\def\@p@@sangle#1{
		\@angletrue
		\edef\@p@sangle{#1} 
}
\def\@p@@ssilent#1{ 
		\@verbosefalse
}
\def\@p@@sprolog#1{\@prologfiletrue\def\@prologfileval{#1}}
\def\@p@@spostlog#1{\@postlogfiletrue\def\@postlogfileval{#1}}
\def\@cs@name#1{\csname #1\endcsname}
\def\@setparms#1=#2,{\@cs@name{@p@@s#1}{#2}}
\def\ps@init@parms{
		\@bbllxfalse \@bbllyfalse
		\@bburxfalse \@bburyfalse
		\@heightfalse \@widthfalse
		\@rheightfalse \@rwidthfalse
		\def\@p@sbbllx{}\def\@p@sbblly{}
		\def\@p@sbburx{}\def\@p@sbbury{}
		\def\@p@sheight{}\def\@p@swidth{}
		\def\@p@srheight{}\def\@p@srwidth{}
		\def\@p@sangle{0}
		\def\@p@sfile{} \def\@p@sbbfile{}
		\def\@p@scost{10}
		\def\@sc{}
		\@prologfilefalse
		\@postlogfilefalse
		\@clipfalse
		\if@noisy
			\@verbosetrue
		\else
			\@verbosefalse
		\fi
}
\def\parse@ps@parms#1{
	 	\@psdo\@psfiga:=#1\do
		   {\expandafter\@setparms\@psfiga,}}
\newif\ifno@bb
\def\bb@missing{
	\if@verbose{
		\ps@typeout{psfig: searching \@p@sbbfile \space  for bounding box}
	}\fi
	\no@bbtrue
	\epsf@getbb{\@p@sbbfile}
        \ifno@bb \else \bb@cull\epsf@llx\epsf@lly\epsf@urx\epsf@ury\fi
}	
\def\bb@cull#1#2#3#4{
	\dimen100=#1 bp\edef\@p@sbbllx{\number\dimen100}
	\dimen100=#2 bp\edef\@p@sbblly{\number\dimen100}
	\dimen100=#3 bp\edef\@p@sbburx{\number\dimen100}
	\dimen100=#4 bp\edef\@p@sbbury{\number\dimen100}
	\no@bbfalse
}
\newdimen\p@intvaluex
\newdimen\p@intvaluey
\def\rotate@#1#2{{\dimen0=#1 sp\dimen1=#2 sp
		  \global\p@intvaluex=\cosine\dimen0
		  \dimen3=\sine\dimen1
		  \global\advance\p@intvaluex by -\dimen3
		  \global\p@intvaluey=\sine\dimen0
		  \dimen3=\cosine\dimen1
		  \global\advance\p@intvaluey by \dimen3
		  }}
\def\compute@bb{
		\no@bbfalse
		\if@bbllx \else \no@bbtrue \fi
		\if@bblly \else \no@bbtrue \fi
		\if@bburx \else \no@bbtrue \fi
		\if@bbury \else \no@bbtrue \fi
		\ifno@bb \bb@missing \fi
		\ifno@bb \ps@typeout{FATAL ERROR: no bb supplied or found}
			\no-bb-error
		\fi
		\count203=\@p@sbburx
		\count204=\@p@sbbury
		\advance\count203 by -\@p@sbbllx
		\advance\count204 by -\@p@sbblly
		\edef\ps@bbw{\number\count203}
		\edef\ps@bbh{\number\count204}
		\if@angle 
			\Sine{\@p@sangle}\Cosine{\@p@sangle}
	        	{\dimen100=\maxdimen\xdef\r@p@sbbllx{\number\dimen100}
					    \xdef\r@p@sbblly{\number\dimen100}
			                    \xdef\r@p@sbburx{-\number\dimen100}
					    \xdef\r@p@sbbury{-\number\dimen100}}
                        \def\minmaxtest{
			   \ifnum\number\p@intvaluex<\r@p@sbbllx
			      \xdef\r@p@sbbllx{\number\p@intvaluex}\fi
			   \ifnum\number\p@intvaluex>\r@p@sbburx
			      \xdef\r@p@sbburx{\number\p@intvaluex}\fi
			   \ifnum\number\p@intvaluey<\r@p@sbblly
			      \xdef\r@p@sbblly{\number\p@intvaluey}\fi
			   \ifnum\number\p@intvaluey>\r@p@sbbury
			      \xdef\r@p@sbbury{\number\p@intvaluey}\fi
			   }
			\rotate@{\@p@sbbllx}{\@p@sbblly}
			\minmaxtest
			\rotate@{\@p@sbbllx}{\@p@sbbury}
			\minmaxtest
			\rotate@{\@p@sbburx}{\@p@sbblly}
			\minmaxtest
			\rotate@{\@p@sbburx}{\@p@sbbury}
			\minmaxtest
			\edef\@p@sbbllx{\r@p@sbbllx}\edef\@p@sbblly{\r@p@sbblly}
			\edef\@p@sbburx{\r@p@sbburx}\edef\@p@sbbury{\r@p@sbbury}
		\fi
		\count203=\@p@sbburx
		\count204=\@p@sbbury
		\advance\count203 by -\@p@sbbllx
		\advance\count204 by -\@p@sbblly
		\edef\@bbw{\number\count203}
		\edef\@bbh{\number\count204}
}
\def\in@hundreds#1#2#3{\count240=#2 \count241=#3
		     \count100=\count240	
		     \divide\count100 by \count241
		     \count101=\count100
		     \multiply\count101 by \count241
		     \advance\count240 by -\count101
		     \multiply\count240 by 10
		     \count101=\count240	
		     \divide\count101 by \count241
		     \count102=\count101
		     \multiply\count102 by \count241
		     \advance\count240 by -\count102
		     \multiply\count240 by 10
		     \count102=\count240	
		     \divide\count102 by \count241
		     \count200=#1\count205=0
		     \count201=\count200
			\multiply\count201 by \count100
		 	\advance\count205 by \count201
		     \count201=\count200
			\divide\count201 by 10
			\multiply\count201 by \count101
			\advance\count205 by \count201
		     \count201=\count200
			\divide\count201 by 100
			\multiply\count201 by \count102
			\advance\count205 by \count201
		     \edef\@result{\number\count205}
}
\def\compute@wfromh{
		\in@hundreds{\@p@sheight}{\@bbw}{\@bbh}
		\edef\@p@swidth{\@result}
}
\def\compute@hfromw{
	        \in@hundreds{\@p@swidth}{\@bbh}{\@bbw}
		\edef\@p@sheight{\@result}
}
\def\compute@handw{
		\if@height 
			\if@width
			\else
				\compute@wfromh
			\fi
		\else 
			\if@width
				\compute@hfromw
			\else
				\edef\@p@sheight{\@bbh}
				\edef\@p@swidth{\@bbw}
			\fi
		\fi
}
\def\compute@resv{
		\if@rheight \else \edef\@p@srheight{\@p@sheight} \fi
		\if@rwidth \else \edef\@p@srwidth{\@p@swidth} \fi
}
\def\compute@sizes{
	\compute@bb
	\if@scalefirst\if@angle
	\if@width
	   \in@hundreds{\@p@swidth}{\@bbw}{\ps@bbw}
	   \edef\@p@swidth{\@result}
	\fi
	\if@height
	   \in@hundreds{\@p@sheight}{\@bbh}{\ps@bbh}
	   \edef\@p@sheight{\@result}
	\fi
	\fi\fi
	\compute@handw
	\compute@resv}
\def\OzTeXSpecials{
	\special{empty.ps /@isp {true} def}
	\special{empty.ps \@p@swidth \space \@p@sheight \space
			\@p@sbbllx \space \@p@sbblly \space
			\@p@sbburx \space \@p@sbbury \space
			startTexFig \space }
	\if@clip{
		\if@verbose{
			\ps@typeout{(clip)}
		}\fi
		\special{empty.ps doclip \space }
	}\fi
	\if@angle{
		\if@verbose{
			\ps@typeout{(rotate)}
		}\fi
		\special {empty.ps \@p@sangle \space rotate \space} 
	}\fi
	\if@prologfile
	    \special{\@prologfileval \space } \fi
	\if@decmpr{
		\if@verbose{
			\ps@typeout{psfig: Compression not available
			in OzTeX version \space }
		}\fi
	}\else{
		\if@verbose{
			\ps@typeout{psfig: including \@p@sfile \space }
		}\fi
		\special{epsf=\ps@predir\@p@sfile \space }
	}\fi
	\if@postlogfile
	    \special{\@postlogfileval \space } \fi
	\special{empty.ps /@isp {false} def}
}
\def\DvipsSpecials{
	\special{ps::[begin] 	\@p@swidth \space \@p@sheight \space
			\@p@sbbllx \space \@p@sbblly \space
			\@p@sbburx \space \@p@sbbury \space
			startTexFig \space }
	\if@clip{
		\if@verbose{
			\ps@typeout{(clip)}
		}\fi
		\special{ps:: doclip \space }
	}\fi
	\if@angle
		\if@verbose{
			\ps@typeout{(clip)}
		}\fi
		\special {ps:: \@p@sangle \space rotate \space} 
	\fi
	\if@prologfile
	    \special{ps: plotfile \@prologfileval \space } \fi
	\if@decmpr{
		\if@verbose{
			\ps@typeout{psfig: including \@p@sfile.Z \space }
		}\fi
		\special{ps: plotfile "`zcat \@p@sfile.Z" \space }
	}\else{
		\if@verbose{
			\ps@typeout{psfig: including \@p@sfile \space }
		}\fi
		\special{ps: plotfile \@p@sfile \space }
	}\fi
	\if@postlogfile
	    \special{ps: plotfile \@postlogfileval \space } \fi
	\special{ps::[end] endTexFig \space }
}
\def\psfig#1{\vbox {
	\ps@init@parms
	\parse@ps@parms{#1}
	\compute@sizes
	\ifnum\@p@scost<\@psdraft{
		\PsfigSpecials 
		\vbox to \@p@srheight sp{
			\hbox to \@p@srwidth sp{
				\hss
			}
		\vss
		}
	}\else{
		\if@draftbox{		
			\hbox{\fbox{\vbox to \@p@srheight sp{
			\vss
			\hbox to \@p@srwidth sp{ \hss 
			 \hss }
			\vss
			}}}
		}\else{
			\vbox to \@p@srheight sp{
			\vss
			\hbox to \@p@srwidth sp{\hss}
			\vss
			}
		}\fi

	}\fi
}}
\psfigRestoreAt
\setDriver
\let\@=\LaTeXAtSign

\newcommand{\stt}{\small\tt}

\begin{opening}
\title{EXPLOSION MODELS, LIGHT CURVES, SPECTRA AND $H_o$}

\author{P. H\"oflich}
\institute{Harvard University, Center for Astrophysics\\
           60 Garden Str.,  Cambridge, MA 02138, USA}
\author{A. Khokhlov}  
\author{J.C. Wheeler} 
\institute{Dept. of Astronomy, U. of Texas, Austin, TX78712, USA}
\author{K. Nomoto} 
\institute{Dept. of Astronomy, U. of Tokyo 113, Japan}
\author{F.K. Thielemann}
\institute{Depart. of Physics, U. Basel, Klingelberg-Str. 62, Switzerland}

\end{opening}

 \runningtitle{EXPLOSION MODELS, LIGHT CURVES, SPECTRA AND $H_o$}

\begin{document}
%
%
\begin{abstract}                 
 From the spectra and light curves it is clear that SNIa events 
are thermonuclear explosions of white dwarfs. However, details 
of the explosion are  highly under debate. Here, we present detailed models                 
which are consistent with respect to the explosion mechanism,
the optical and infrared light curves (LC), and the spectral evolution. This leaves the
description of the burning front and the structure of the white dwarf as
the only free parameters. The explosions are calculated using  one-dimensional
Lagrangian codes including nuclear networks. Subsequently, optical and IR-LCs are
constructed. Detailed NLTE-spectra are computed for several instants of time using
the density, chemical and luminosity structure resulting from the LCs.
 The general methods and critical tests are presented (sect. 2).
 
 Different models for the thermonuclear explosion are discussed including
detonations, deflagrations, delayed detonations, pulsating delayed detonations (PDD)
 and helium detonations (sect.3).
Comparisons between theoretical and observed LCs and spectra provide
an insight into details of the explosion and nature of the progenitor stars (sect. 4 \& 5).
 We try to answer several related questions.
 Are subluminous SNe~Ia a group different from `normal' SN~Ia (sect. 5)?
  Can we understand observed properties of the LCs and spectra
(sect. 4)?
 What do we learn about the progenitor evolution and its metallicity
(sect. 3, Figs. 4,5)?
 Do successful SN~Ia models depend on the type of the host galaxy (Table 2)?
 
 Using both the spectral and LC  information,
 theoretical models allow for a determination of the Hubble constant independent 
from `local' distance indicators such as $\delta $ Cephei stars.
 $H_o$ is found to be $67\pm~9~km~s^{-1}Mpc^{-1}$ and, from SN1988U, $q_o$ equals
$0.7 \pm 1.$ within 95 \% confidence levels.
%
\end{abstract}

\section{Introduction}
During the last few years it became evident that Type Ia supernovae
are a less homogeneous class than previously believed (e.g. Barbon et al., 1990,
Pskovskii 1970,       Phillips et al. 1987).
 In particular,
the observation of subluminous SN1991bg 
 (Fillipenko et al. 1992, Leibundgut et al. 1993)
 raised questions on the SN-rate   and whether we have missed a huge subgroup. 
The possible impact on our understanding of supernovae statistics and,
consequently, the chemical evolution of galaxies must be noted.

 It is generally accepted that SNe~Ia are
  thermonuclear explosions of carbon-oxygen white dwarfs (WD)
(Hoyle \& Fowler 1960). However, details of the scenario are still under debate. 
 For discussions of various theoretical aspects see
e.g. Wheeler \& Harkness (1990), Woosley \& Weaver (1994), Canal (1995),
Nomoto et al. (1995), Wheeler et al. (1995).
 
 What we observe as a supernovae event is not the explosion itself but the light 
emitted from a rapidly expanding envelope produced by the stellar explosion. As 
the photosphere recedes, deeper layers of the ejecta become visible. A detailed 
analysis of the LCs and spectra gives us the opportunity to determine the
density, velocity and composition structure of the ejecta.
                           A successful application of these
constraints, however, requires both accurate early LC and spectral
observations and
detailed theoretical  models.  While SN~Ia LC data have enormously
increased, until
recently (Harkness 1991; H\"oflich et al. 1991),
 models are often hampered by inadequate physical assumptions
like a constant opacity, crude gamma-ray deposition schemes and a simplified
treatment of the ionization balance, neglect of line scattering 
and by the use of the diffusion approximation 
(e.g. Livne \& Arnett 1995). Based on our detailed models which overcome
the approximations just mentioned,
 we have investigated the validity and influence of the physical assumptions 
made in LC and spectral calculations and  estimated the 
accuracy of our models.
The importance  
of a consistent treatment of the explosion mechanism, LCs and
spectra became evident.
 Some of the tests and the basic outline of our approach
is given in section 2. In section 3, the basic explosion mechanism are discussed.
After investigating the general properties of our LCs and spectra, individual
comparisons with observations are presented. Finally, we address the question on
$H_o$ and $q_o$.

\section{Brief Description of the Numerical Methods}

\subsection { Hydrodynamics}
 
 The explosions are calculated using  one-dimensional
Lagrangian hydro with artificial viscosity (Khokhlov, 1991ab) 
and  radiation-hydro codes including nuclear networks  (H\"oflich \& Khokhlov
 1995).
 The latter code is based on our LC code that solves the hydrodynamical equations explicitly
by the piecewise parabolic method (Collela and Woodward 1984) and includes the solution of the radiation transport
implicitly via the moment equations, expansion opacities,
 and a  detailed equation of state. Typically,  300 to  500 depth points are used.
 Radiation transport has been included   
to provide a smoother transition from the hydrodynamical explosion to the phase
of free expansion.  We omit $\gamma$-ray transport during the hydrodynamical phase  because of the 
high optical depth of the inner layers.
 Nuclear burning is taken into account using Thielemann's  network
 (Thielemann, Nomoto \&   Hashimoto 1994
 and references therein). 
 During the hydro, 
 an $\alpha$-network of 20 isotopes is considered to properly 
describe the energy release. Based on a network of 216
isotopes, the final chemical
structure is calculated  by postprocessing the hydrodynamical model.
 The accuracy of the energy release has been found
to be about 1 to 3 \% in the reduced network compared to about 15 \% for
the equation net work used previously (Khokhlov, 1991) which, in general, overestimate
the expansion velocity accordingly.
 \begin{figure}
\psfig{figure=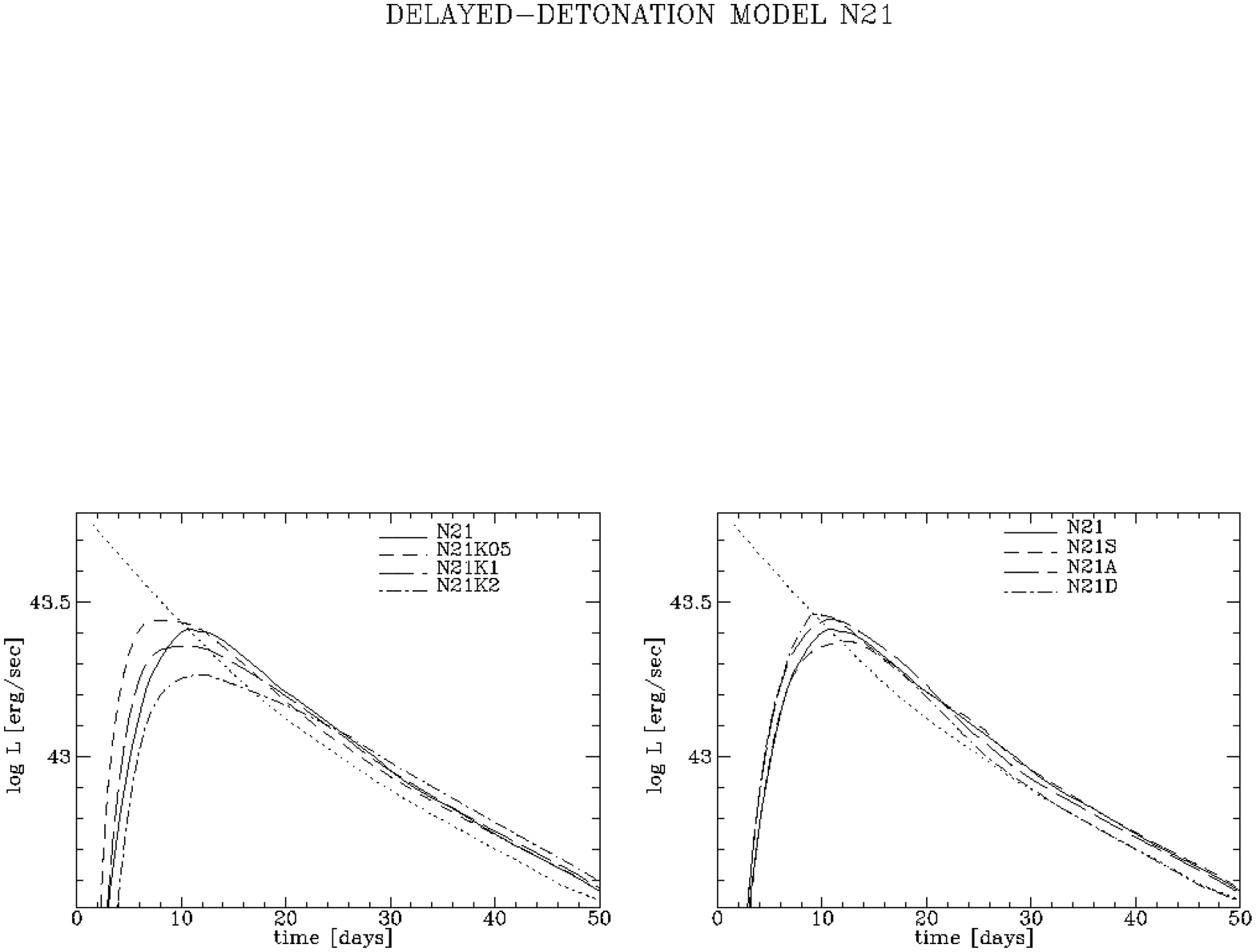,width=12.6cm,rwidth=4.5cm,angle=270}
 \caption{
Bolometric LC of the delayed detonation model
N21 for various assumptions (N21: all effects are included as described in
the text; N21K2, N21K1 \& N21K05: 
 $\kappa _R = \kappa _P = const $ with
 0.20, 0.1 and 0.05 $~cm^2 /g$, respectively;
 N21S: $\kappa _P = \kappa _R$, N21A; extinction = true absorption;
N21D: diffusion approximation). In addition the energy release
due to the radioactive decay of $^{56}Ni $ and $^{56}Co $ is shown (dotted line).
 If constant opacities are assumed, explosion models cannot be tested because
the  strong temperature dependence of the opacity for $T \leq 20000 K $ is
neglected. This effect determines the shape of LCs near maximum light.
For reasonable variations in $\kappa $ (Shigeyama et al. 1993),
 the time of
maximum light is shifted by up to 5~days and the maximum brightness
may be off by about 50\% . The shape of the LCs differ
 strongly. If the
 temperature dependence of the opacity is included,
the spread in the results
shrinks to about 20 \% both with respect to the rise time and absolute
flux. The slope becomes comparable within 10\% if the radiation transport equation 
is solved rather than assuming the diffusion approximation.
 }
 \end{figure}
 
\subsection { Light Curves}
 
Based on the explosion models,  further
hydrodynamical evolution, and  bolometric and monochromatic light
curves are calculated using a scheme recently developed 
 and widely applied to  SN Ia (H\"oflich et al. 1995 and references therein).
 In principle, it is the same code as that described above, but 
 nuclear burning is neglected and
 $\gamma $ ray transport is included via a Monte Carlo scheme.
In order to allow for a more consistent treatment of scattering, we
solve both the (two lowest) time-dependent radiation moment equations for the radiation
energy and the radiation flux, and a total energy equation 
(Mihalas 1978). The Eddington factors are obtained by formal 
integration of the radiation transport equation.
 The opacities have been calculated under the assumption
of local thermodynamical equilibrium. This is a reasonable approximation 
 since diffusion time scales are governed by layers
of large optical depth.
 \begin{figure}
\psfig{figure=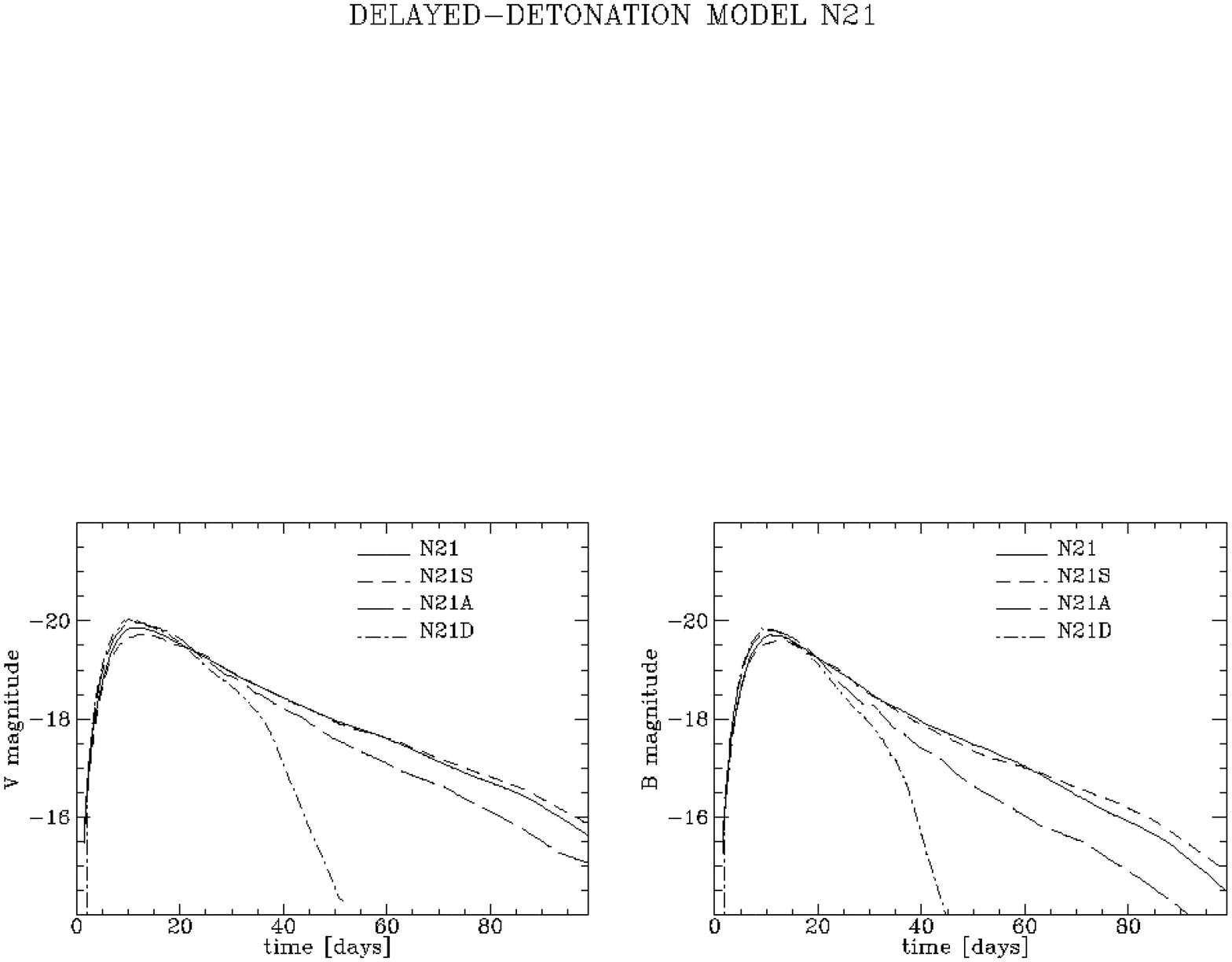,width=12.6cm,rwidth=4.5cm,clip=,angle=270}
 \caption{
 B and V LCs of              
N21 for various assumptions (see Fig. 1). Apparently, both
the diffusion approximation (N21D) and the neglection of line
scattering (extinction = true absorption, N21A) are inadequate. The  
the local cooling of the radiaton field is strongly overestimated 
since, in SNeIa, scattering exceeds the true absorption by a factor
of 100 to 10000 (H\"oflich et al. 1993, H\"oflich 1995). In comparison,
the corresponding error by setting $\kappa _R * \alpha = \kappa _P$
in the energy equations results in errors of about $0.1~m$ because 
the thermal coupling differs by a much smaller
 factor ($\approx 5~ ...~ 10 $). The surprisingly small sensitivity of the 
resulting LCs can be understood in terms of the energy conservation. Namely,
during early times, the energy balance is determined by radioactive decay and the 
adiabatic expansion if the opacity exceeds a certain limit.
 }
\end{figure}
 The Planck and the Rossland mean
of the absorption opacity and extinction, respectively, have been calculated based on
an approach similar to Karp et al. (1977) 
 but transformed to the comoving frame. Relativistic terms are neglected.
As in any Sobolev approach, it is assumed that the 
resonance region of lines is small. Therefore,
 the mean free path of a photon along a certain ray is given 
by the product of inverse absorption probabilities weighted with the absorption probabilty
of transitions being closer to the comoving frame wavelength. The expansion
effect enters both the Planck and the Rosseland mean but the former in second order, only, because
the Planck mean is a linear average of $\kappa _\nu $. The main problem with the Planck opacity
is related to line scattering. In SNeIa, scattering 
exceeds true absorption by several orders of magnitude. Therefore, scattering must    
be included.  Since 1993, 
we have used an equivalent-two level approach calibrated by detailed NLTE-calculations 
(see below, H\"oflich 1995). We found that the thermalization process is governed by  
redistribution (i.e. collisions) of energy within the atomic fine structure levels 
 and not by the scattering in about a dozen
UV lines (Pinto, talk at this meeting). The former effect is dominant since collisional cross
sections depend exponentially on the energy difference between levels.
In fact, the two-level approach 
underestimates the size of the thermalization by  several orders of magnitude as it is well
known in  the field of stellar atmospheres (e.g Mihalas 1978, Ayres 1989).
 For details, see H\"oflich (1995).

Monochromatic colors $B$, $V$, $R$, and $I$ 
are calculated using 100 discrete wavelength bands and formal integration along rays.
 For several models,
detailed NLTE-spectra have been constructed (H\"oflich, 1995).
The colors based on the NLTE atmospheres and LC calculations have been compared.
Typically, the disagreements between these two techniques in V, B, R, and I
 are below $0.05 $ to $0.1^m$ 
near maximum light and remain between 0.2 to 0.4$^m$ at late time. 
The error is  small because
most lines are scattering dominated. Consequently, the flux is 
redistributed within a wavelength range corresponding to the Doppler 
shift produced by the photospheric expanson. However, this frequency shift is  small compared
to the bandwidth of the UBVRI-filter. With increasing time,   
the forbidden lines dominate the flux and, consequently, the error in our
monochromatic colors increases.
Another uncertainty when comparing with observations is due to the filter functions. Here, we use
transmission functions given by Bessell (1990).
 
 In light of the uncertainties, we have compared our opacities with those based on approximations
by Jeffery  (1995). We found good agreement. 
 Moreover, the sensitivity of the resulting LCs
variations in the uncertain quantities have been studied in great detail (H\"oflich et al. 1993,
Khokhlov et al. 1994). Some of the results are shown in Figs. 1 and 2.
   Other, independent tests are provided by the 
comparisons with observations. We find that our
bolometric correction is consistent with the observation of SN1992A,
and the distances predicted
by our models agree with
those based on  $\delta $-Ceph stars in 
IC4182, NGC5253 and 4536 (H\"oflich \& Khokhlov 1995). The former test probes the overall
energy distribution, whereas the latter test is sensitive to errors in the absolute, 
 monochromatic fluxes.

\subsection {Synthetic spectra }

\bigskip\noindent
 A modified version of our  code for Nlte Extended ATmospheres (NEAT) is used.
 For details  see   H\"oflich (1990, 1995) and references therein.

  Stationarity is assumed where the density, chemical profiles and the luminosity
 as a function of the radial distance r are given by results of the hydrodynamical explosion and of the
 LC calculations including  the Monte-Carlo scheme for the $\gamma $-ray
  transport (Fig. 3).

\begin{figure}
\psfig{figure=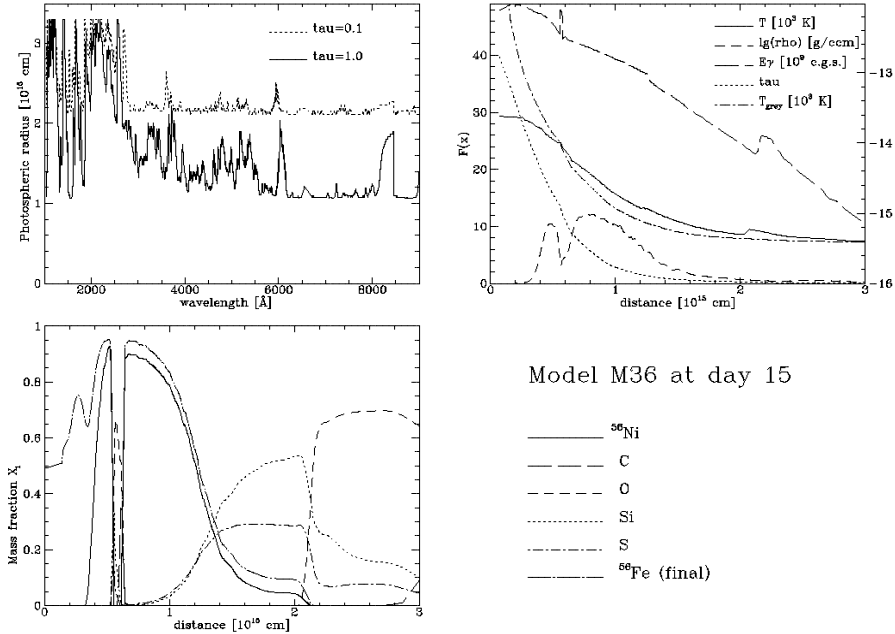,width=12.6cm,rwidth=9.5cm,angle=270}
\caption{ Various quantities for M36 15 days after the explosion.
     Distances  are given as a function of wavelength at which 
the monochromatic optical depth reaches 0.1 and 1, respectively
 (upper left),
          Temperature T, density $\rho $, energy deposition due to radioactive
decay $E_{\gamma} $ and Rosseland optical depth are given as a function of distance
 and, for comparison, $T_{grey}$ for the grey extended atmosphere (upper right),
and chemical profile are presented for  the most abundant elements (lower left).
 No well defined photosphere exists.
The need for  spectral analyses consistent with respect
 to the explosion mechanism,
$\gamma-$ray transport and LC calculation is obvious.
 Already at maximum light, the chemical distribution of elements does not follow
the density profile but shows large individual variations in the 
line forming region, i.e. between 1 and 2 $10^{15}~cm$. The smallest expansion velocity
indicated by line shifts are predominantly determined by the chemical profiles and not
by the density.  The energy deposition
and excitation due to $\gamma $-rays becomes important within the photosphere.
Consequently, the luminosity cannot be assumed to be depth independent
for the  construction of synthetic spectra.
 }
 \end{figure}

  The radiation transport equation is solved in the comoving frame
including relativistic terms.
For the determination of the first moment of the intensity, we solve
the radiation transport equation for strong lines including `quasi' continua
 (weak lines are treated in a Sololev approximation, see below)
 in the comoving frame, i.e.            
$$
 Op~I = \chi (S  -  I)
$$
$$
I = I (\mu, r, \nu);
~~~~~~~~~S = S ( r, \nu);~~~~~~~~
\chi = \chi ( r, \nu) 
$$
($\mu $: cosine between the radial direction and I; r : radial distance;
v(r) : radial velocity; S: source function; $\chi $: absorption
coefficient; $\nu $: frequency).
       The operator $Op$ can be written as follows (Castor, 1974):
 
$$
Op = \Biggl\lbrack\mu +{v(r) \over c}\Biggr\rbrack
 {\partial \over \partial r}\\
+\Biggl\lbrack {(1 - \mu^2) \over r}
 \Bigl(1+ {\mu v(r) \over c} (1-\beta(r))\Bigr)
\Biggr\rbrack      {\partial \over \partial \mu} $$
$$-\Biggl\lbrack {v(r) \nu \over c \cdot r} \Bigl(1 -
\mu^2 + \mu^2 \beta(r)\Bigr) \Biggr\rbrack
{\partial \over \partial \nu} \\
+ \Biggl\lbrack  {3 v(r) \over c \cdot r}
\biggl( 1-\mu^2+ \beta(r) \mu^2 \biggr)
 \Biggr\rbrack\\
$$
 
with
$$
\beta : = {d~ln~v(r) \over d~ln~ r}. 
$$
 The term in front of $  {\partial \over \partial \nu} $ can be
interpreted as the classical Doppler shift. The second terms of the
partial derivatives with respect    
to r, $\mu $ and the logarithmic derivative of v(r) 
  correspond to the advection and aberration effects.
 For practical purpose, we solve the non relativistic transport equation using a
Rybicki scheme (Mihalas et al., 1975, 1976).
 Overlapping lines in the comoving frame
 cannot be treated because the system is
solved as a boundary problem in the frequency space which, from the concept, disallows
a propagation of information toward higher/lower frequencies for expanding/collapsing 
envelopes. The same limitation should also apply to Nugent et al. (1995a) who use the same
method.
However, the problem of overlapping lines can be and has been solved (H\"oflich 1990)
 in full analogy with 
the problem of partial redistribution using a Fourtrier-like scheme (Mihalas et al, 1976b).
Although our code can deal with partial redistribution, this effect is
neglected to save CPU-time.
 A technical problem is the fact that
each line must be sampled by  at least 10  frequency points to provide a sufficient accurate
representation of the $\delta /\delta \nu$-term (Mihalas et al. 1975). Therefore, we must limit 
this approach to  $\approx 2000 $ lines.    

 Blocking by lines other than the strong ones  is included in 
   a `quasi' continuum approximation, i.e. the frequency derivative terms in the radiation 
transport equation is included in the narrow line limit  to calculate 
the probability for photons  to pass a radial sub-shell along a given direction $\mu $ 
(Castor, 1974,  Abbot and Lucy  1985, H\"oflich, 1990).
 Note that the information on the exact location of the interaction within a subshell gets lost 
translating  into an uncertainty in the wavelength location of $\approx 10^{-3}$.

   The statistical equations are solved consistently  with the radiation transport
 equation to determine the non-LTE occupation numbers using both an accelerated lambda
iteration (see  Olson  et al.  1986) 
         and an  equivalent-two level approach for               
 transitions from the ground state which provides an efficient way to take the non-thermal
fraction of the source function into account during the radiation transport and, effectively,
accelerate both the convergence rate and the stability of the system. A comparison of the         
    explicit with the implicit source functions provides a sensitive tool to test for  
convergence of the system of rate and radiation transport equations.
 
 Excitation by     gamma rays
is included. Detailed atomic models are used for up to three most abundant ionization stages
of several elements, i.e. (He), C, O, Ne, Na, Mg, Si, S, Ca, Fe taking into account
20 to 30 levels and  80 - 150 transitions in the main ionization stage. Here, we use
detailed term-schemes for C~II, O~II, Ne I, Na~I, Mg~II, Si~II, S~II, Ca~II and Fe~II or Fe~III.
 The corresponding lower and upper ions are represented by the ground states.
 The energy levels and cross sections of  bound-bound transitions are taken from
 Kurucz (1993).
 In addition to our NLTE-transitions, a total of $\approx 300,000$ lines out of
a list of 31,000,000 (Kurucz, 1993) are included for the radiation transport.                  
 For these lines, we assume
LTE population numbers inside each ion. To calculate the ionization balance,
 excitation by     hard radiation is taken into
 account.   LTE-line scattering is included using  an equivalent-two-level
approach calibrated by our NLTE-elements.

\section{Hydrodynamical Models}
 
\subsection{Explosions of Massive White Dwarfs}
 
 A   first group consists of massive carbon-oxygen white
dwarfs (WDs) with a mass close to the Chandrasekhar mass which accrete mass
through Roche-lobe overflow from an evolved companion star (Nomoto \&
Sugimoto 1977; Nomoto 1982).  The explosion is
triggered by compressional heating. 
                                 The key question is 
how the flame propagates through the white
dwarf. Several models of SNeIa have been proposed in the
past, including detonation (Arnett 1969; Hansen \& Wheeler 1969), 
deflagration (e.g. Nomoto, Sugimoto \&
Neo 1976) and the delayed detonation model,                     
which assumes that the flame starts as a deflagration and turns into a 
detonation later on (Khokhlov 1991, Woosley \& Weaver 1995, Yamaoka et al.  
1992).
\begin{table}
\begin{center}
\caption{
Some quantities (see text)
 are given for detonations (DET1/DET2, asterix), deflagrations
(W7, DF1, DF1mix, circles), delayed detonations (N21, N32, M35-M312, DD13c - 27c, bullets),
pulsating delayed detonations (PDD1a-9, black triangles),
envelope models (DET2env2/4, open triangles), and helium detonations (HeD2-12, open asterix).
 The C/O ratio has been assumed to be 1:1 unless otherwise quoted in brackets
after the name. The initial metallicity for $Z\geq 20$ is assumed to be solar,
but for DD24c, DD25c, DD26c and DD27c for which 1/3, 3, 0.1 and 10 
times solar abundances are used, respectively. Otherwise, these models are 
identical to DD21c.
For the helium detonations and envelope models,
the mass is of the C/O core, and of the He-layers  or of the extended CO-envelope,
respectively, are given separately. In HeD7, no central C-ignition is triggered.
}
\begin{tabular}{llllllll}
\hline 
  Model~~~~~ &  $M_\star$ ~~~~~& $\rho_c$ ~~~~~~~~~ & $\alpha$ ~~~~~&
  $\rho_{tr}$~~~~~~~~~~~~ &
 $E_{kin}$~~~~~  & $M_{Ni}$~~~~~  \\
& $[ M_\odot ]$ & $[10^9 c.g.s]$~ &  &   $[ 10^7 c.g.s]$~ & $[   10^{51} erg    ]~$ & $[ M_\odot ]$ &     \\
\hline                     
  DET1   &   1.4  &  3.5   &   ---  &  ---  &  1.75  &
                          0.92  \\
DF1    &   1.4  &  3.5   &  0.30  &  ---  &  1.10  &
                          0.50  \\
W7     &  1.4  &  2.0   &  n.a.  &  ---  &  1.30  &
                          0.59  \\
N21     &   1.4  &  3.5   &  0.03  &  5.0  &  1.63  &
                           0.83  \\
N32     &   1.4  &  3.5   &  0.03  &  2.6  &  1.52  &
                           0.56  \\
M35     &  1.4  &  2.8   &  0.03  &  3.0  &  1.56  &
                           0.67  \\
M36     &   1.4  &  2.8   &  0.03  &  2.4  &  1.52  &
                           0.60  \\
M37     &   1.4  &  2.8   &  0.03  &  2.0  &  1.49  &
                           0.51  \\
M38     &   1.4  &  2.8   &  0.03  &  1.7  &  1.44  &
                           0.43  \\
M39     &   1.4  &  2.8   &  0.03  &  1.4  &  1.38  &
                           0.34  \\
M312    &   1.4  &  2.8   &  0.03  &  1.0  &  1.35  &
                           0.20  \\
DD13c (1:1)  & 1.4  &  2.6   &  0.03  &  3.0  &  1.36  &
                           0.79  \\
DD14c (1:2)  &  1.4  &  2.6   &  0.03  &  3.0  &  1.21  &
                           0.79  \\
DD15c (2:3)   &  1.4  &  2.6   &  0.03  &  3.0  &  1.28  &
                           0.79  \\
DD21c (1:1)   &  1.4  &  2.6   &  0.03  &  2.7  &  1.32  &
                           0.69  \\
DD23c (2:3)   &  1.4  &  2.6   &  0.03  &  2.7  &  1.18  &
                           0.59  \\
PDD3    &    1.4  &  2.1   &  0.04  &  2.0  &  1.37  &
                           0.49  \\
PDD5    &                 1.4  &  2.7   &  0.03  &  0.76 &  1.23  &
                           0.12  \\
PDD8    &                 1.4  &  2.7   &  0.03  &  0.85 &  1.30  &
                           0.18  \\
PDD7    &                 1.4  &  2.7   &  0.03  &  1.1  &  1.40  &
                           0.36  \\
PDD9    &                 1.4  &  2.7   &  0.03  &  1.7  &  1.49  &
                           0.66  \\
PDD6    &                 1.4  &  2.7   &  0.03  &  2.2  &  1.49  &
                           0.56  \\
PDD1a   &                 1.4  &  2.4   &  0.03  &  2.3  &  1.65  &
                           0.61  \\
PDD1c   &                 1.4  &  2.4   &  0.03  &  0.71 &  0.47  &
                           0.10  \\
HeD2    &                 0.6+0.22  &  .013   &  ---   &  ---  &  0.94  &
                           0.43  &   \\
HeD4    &                 1.0+0.18  &  .150   &  ---   &  ---  &  1.50  &
                           1.07  &   \\
HeD6    &                 0.6+0.172 &  .0091  &  ---   &  ---  &   0.72 &
                           0.252 &   \\
HeD7    &                 0.6+0.14  &  .0089  &  ---   &  ---  &   ---- &
                           ----  \\
HeD8    &                 0.8+0.16  &  .025  &  ---   &  ---  &   1.08 &
                           0.526 \\
HeD10   &                 0.8+0.22  &  .036  &  ---   &  ---  &   1.24 &
                           0.75  \\
HeD11   &                 0.9+0.16  &  .061  &  ---   &  ---  &   1.37 &
                           0.87  \\
HeD12   &                 0.9+0.22  &  .083  &  ---   &  ---  &   1.45 &
                           0.92  \\
 DET2     &               1.2        &  0.04  &  ---  &  ---  &
                              1.52  &  0.63  \\
DET2env2 &               1.2 + 0.2  &  0.04  &  ---  &  ---  &
                             1.52  &  0.63  \\
DET2env4 &               1.2 + 0.4  &  0.04  &  ---  &  ---  &
                             1.52  &  0.63  \\
 DET2env6 &               1.2 + 0.6  &  0.04  &  ---  &  ---  &
                              1.52  &  0.63  \\
\hline         
\end{tabular}
\end{center}
\end{table}      
  
 \begin{figure}
\psfig{figure=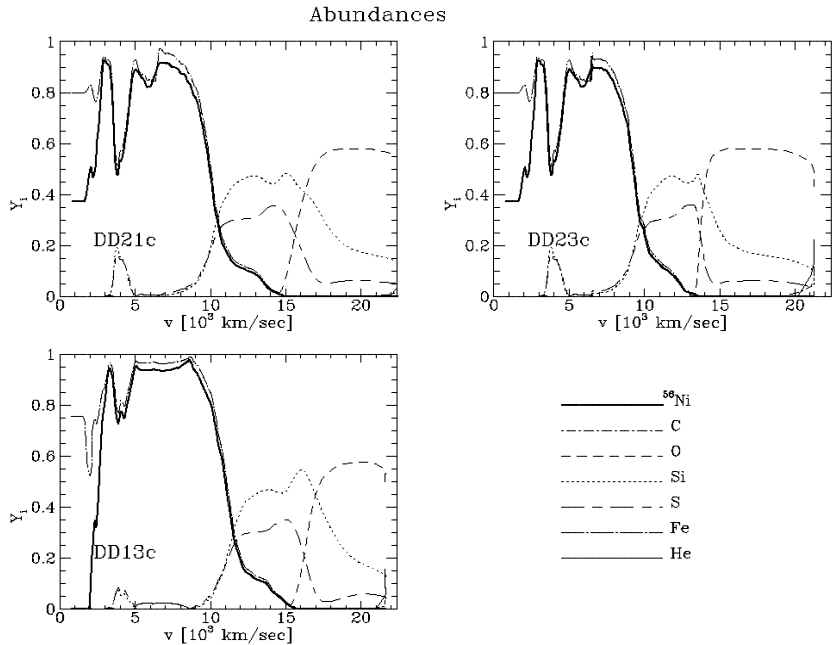,width=12.6cm,rwidth=8.7cm,angle=270}
 \caption{
Abundances as a function of the expansion velocity for
three delayed detonation models (see table 1). Since the burning time
scales to NSE or partial NSE are shorter than the hydrodynamical time
scales for all but the very outer layers, the final products depend 
mainly on the density at which burning takes place. 
With decreasing
transition density, lesser $^{56}Ni$ is produced and the intermediate
mass elements expand at lower velocities
 because the later  
transition to a detonation allows for a longer 
pre-expansion of the  outer layers (DD21c vs. DD13c). Similarily, 
with increasing C/O ratio in the progenitor, the specific energy
release during the  nuclear burning is reduced (DD23c vs. DD21c) and
the transition density at the burning front reached later in time, 
resulting in a larger preexpansion of the outer layers. This
may  allow to determine the main sequence mass of the progenitor. }
 \end{figure}
\begin{figure}
\psfig{figure=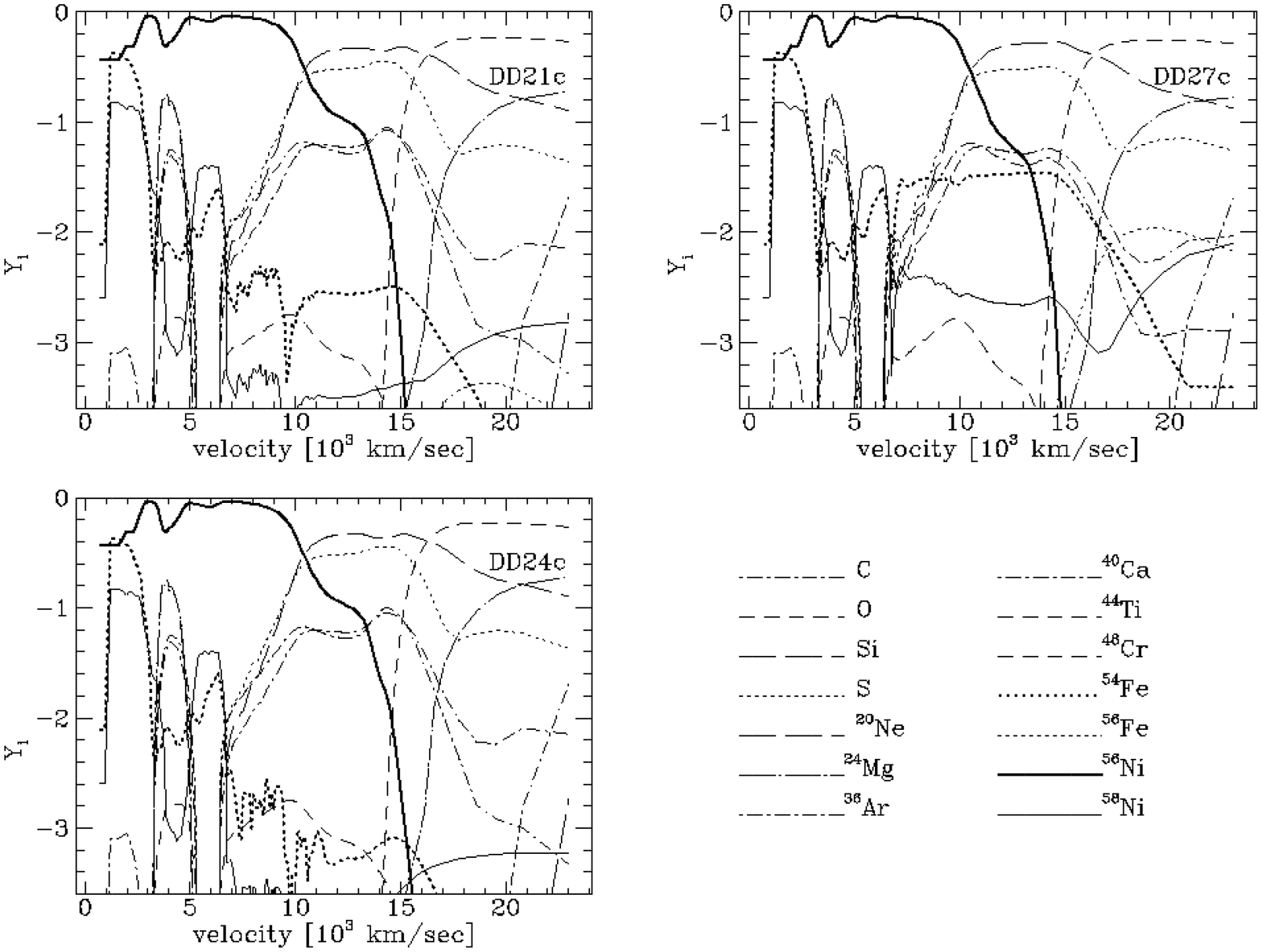,width=12.6cm,rwidth=8.6cm,angle=270}
\caption{
Abundances for
three delayed detonation models (DD21c, DD24c \& DD27c)  
with identical parameters but different initial
metallicity (Z $\geq 20$)
corresponding to 1, 1/3  and 10 times solar. The overall 
density, velocity and chemical structure remains uneffected but
the isotopic composition changes drastically.
 In particular, the final $Fe$ abundance differ by more than 
two orders of magnitude in the outer layers and may be as high as
7 \%  which clearly shows up in synthetic
spectra. Whether this provides a natural explanation of the spectral 
peculiarities of  SN1991T is under investigation.}
\end{figure}
Our sample includes
     detonations (DET1/2), deflagrations (W7, DF1, DF1mix),
 delayed detonations (N21/32, M35-39; DD13-27) and 
 pulsating delayed detonations (PDD1-9).
 The deflagration speed is parameterized as $D_{def} = \alpha a_s$, where $a_s$ is the local sound velocity
ahead of the flame and $\alpha $ is a free parameter. The speed of the detonation wave is given by the
sound-speed behind the front. For delayed detonation models, the transition to a detonation is given 
by another free parameter $\rho _{tr}$. When the density ahead of the deflagration front
reaches $\rho_{tr}$, the transition  to a detonation is forced by increasing $\alpha $ to 0.5 over 
5 time steps bringing the speed well above the Chapman-Jouguet theshold
for steady deflagration.
 For pulsating delayed detonation models,
 the initial phase of burning fails to release sufficient energy to disrupt the WD.
During the subsequent contraction phase, compression of the mixed layer of 
products of burning and C/O formed 
at the dead deflagration front would give rise to a detonation via
compression and  spontaneous ignition (Khokhlov 1991).
 In this
scenario, $\rho_{tr} $ represents  the density at which 
the detonation is initiated after the burning front dies out.
 Besides the description of the burning front, 
 the central density of the WD at the time of the explosion is another free parameter. For white
dwarfs close to the Chandrasekhar limit, it  depends sensitively on the   
chemistry and the accretion rate $\dot M $ at the time of the explosion.
 In our recent study on SN1994D,           
evidence was found that the models with a somewhat smaller 
central density provide better agreement with 
both the observed spectra and LCs.
Therefore,   the grid of  PDD models has been extended.
 
\subsection{Merging White Dwarfs}
 The second group of progenitor
models consists of two low-mass white dwarfs in a close orbit which
decays due to the emission of gravitational radiation and this, eventually,
leads to the merging of the two WDs
 (e.g. Iben \& Tutukov 1984).
After the initial merging process,
one low density WD          is surrounded by an extended envelope
 (Hachisu, Eriguchi \& Nomoto 1986ab,  Benz,   Thielemann \& Hills     1989).
 This scenario is mimicked by our envelope models DET2env2...6 
 in which we consider the detonation in a
low mass WD          surrounded by a compact envelope between 0.2 and 0.6 $M_\odot $.

\subsection{Explosions of Sub-Chandrasekhar Mass White Dwarfs}
      Another class of models -- double detonation of a C/O-WD
triggered by detonation of helium layer in low-mass WDs         
 -- was explored by Nomoto (1980),
Woosley \& Weaver (1980), and most recently by Woosley and Weaver (1994,
hereafter WW94) and H\"oflich \& Khokhlov (1995).
This scenario was also suggested for the explanation of 
subluminous Type~Ia (WW94). Note that the explosion of a low mass WD 
was also suggested by Ruiz-Lapuente et al. (1993) to explain subluminous SNe~Ia
but the mechanism for triggering of the central carbon detonation was not considered.

 For ease of comparison,
 we have  used parameters  close  to those suggested in WW94.
 To prevent repetition,
we refer to the latter work for a detailed discussion of this 
 class of models.
 Helium detonations show a qualitatively different structure in comparison to all models with a 
 Chandrasekhar mass WD           
. The intermediate mass elements
are sandwiched by  Ni and He/Ni rich layers at the inner and outer regions, respectively. 
 Generally,  the density   smoothly decreases  with
mass because partial burning produces almost the same amount of kinetic energy as  the total
burning, but a moderate  shell-like structure is formed just below the former Helium layers.
 Observationally,  a distinguishing feature of this scenario  
 is the presence of Helium and Ni with expansion velocities above 11,000 to
14,000 $km~ s^{-1}$.
Typically  0.07 to 0.13 $M_\odot$ of Ni are produced in the outer layers, mainly depending on the mass of the
Helium shell. Recently,
 Benz (1995, private communication) suggested that the explosion of the C/O core may be triggered
directly by the in-going shock front if Helium is ignited somewhat above the He/C-O interface.
 This, however, should hardly change the chemical structure because the flame propagates
as a detonation.
%
 
\begin{figure}
\psfig{figure=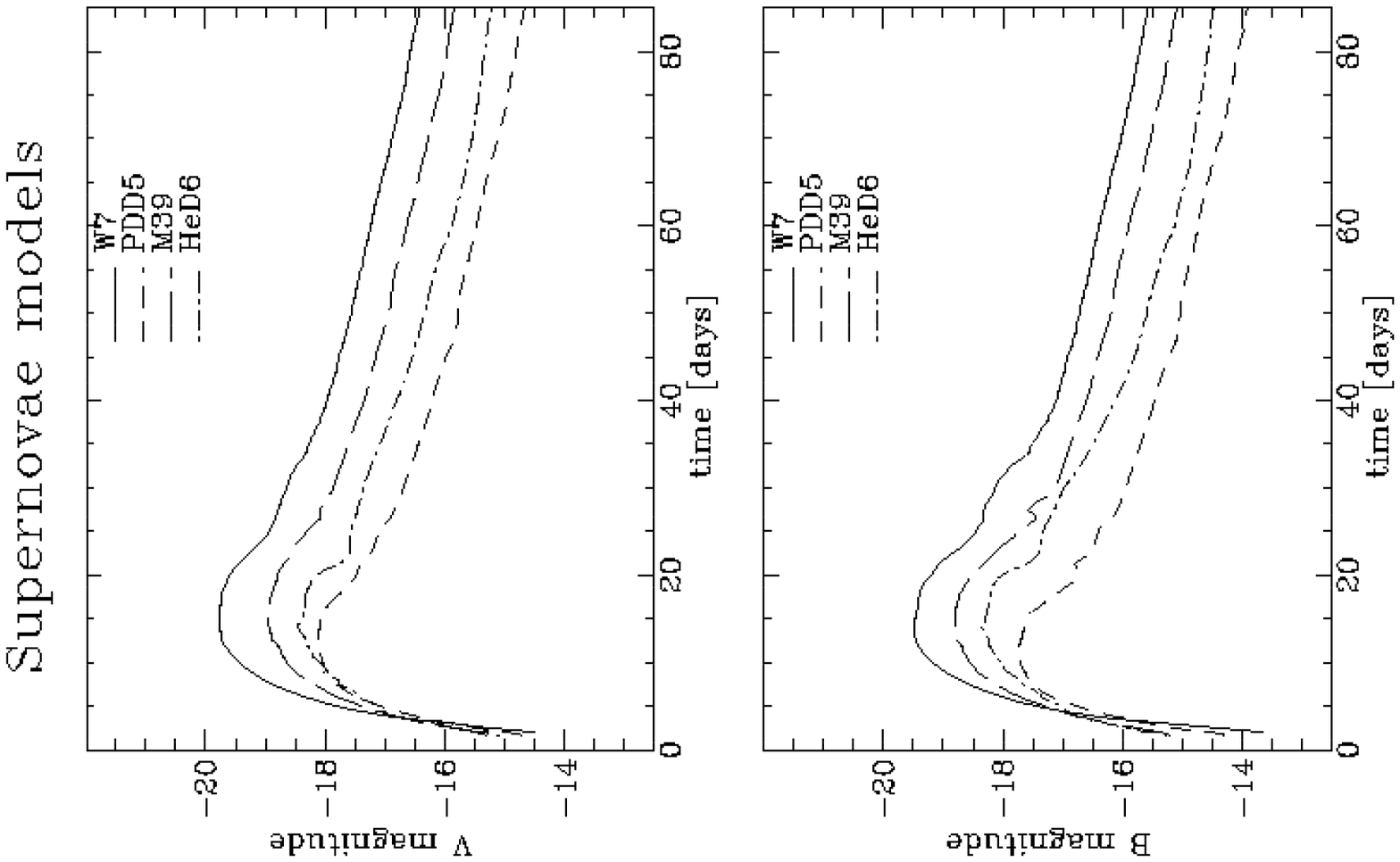,width=12.4cm,rwidth=4.4cm,clip=,angle=270}
\vskip -4.0cm
\psfig{figure=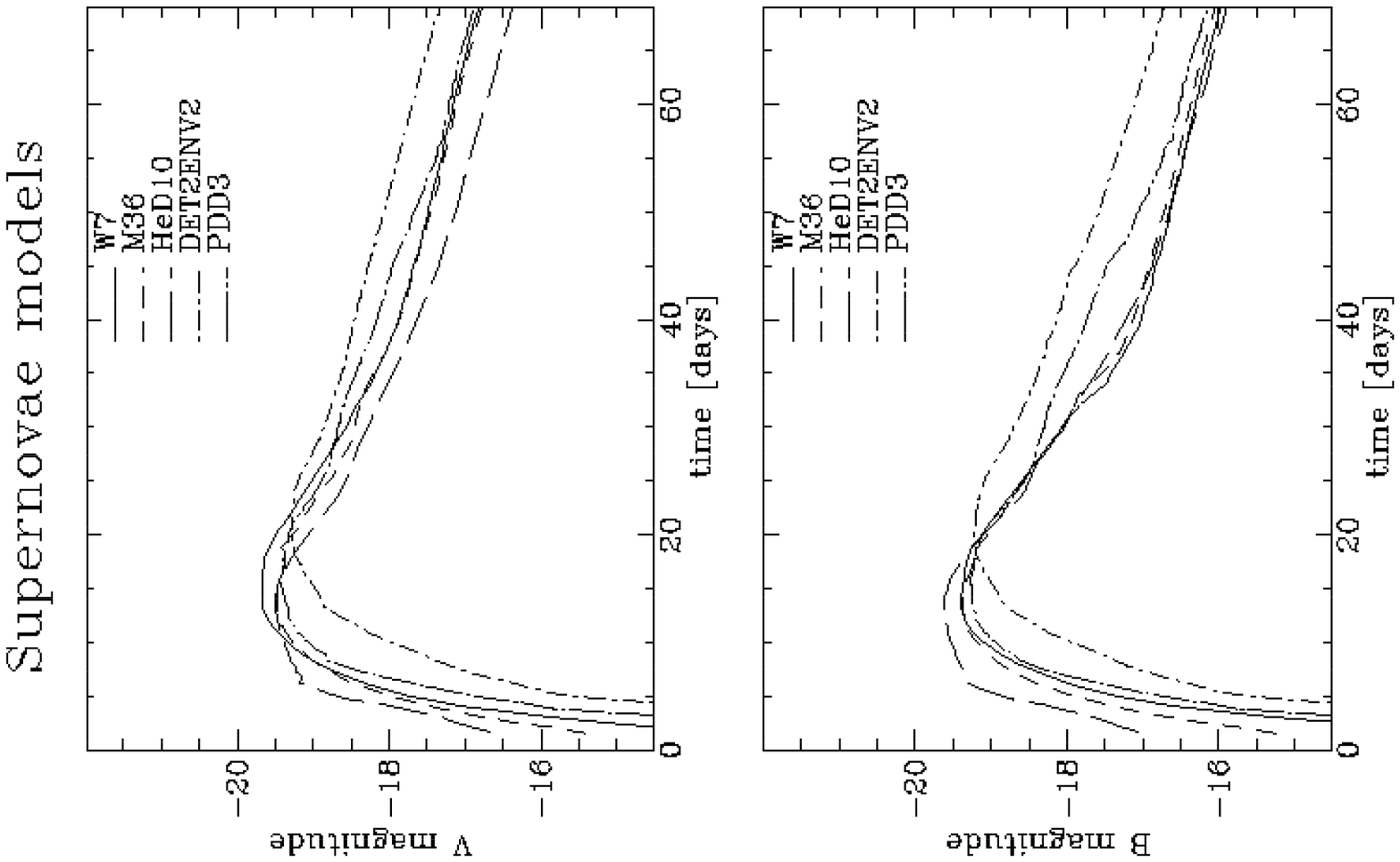,width=6.2cm,rwidth=4.4cm,clip=,angle=270}
\caption{
Monochromatic V LCs of normal bright supernovae of deflagration
(W7), classical delayed detonation (M36), Helium detonation (HeD10), envelope
 (DET2env2), pulsating delayed detonation (PDD3) models, and of subluminous SNeIa
 classical delayed detonation (M39), Helium detonation (HeD6), pulsating delayed detonations(PDD5).
}
\end{figure}
\begin{figure}
\psfig{figure=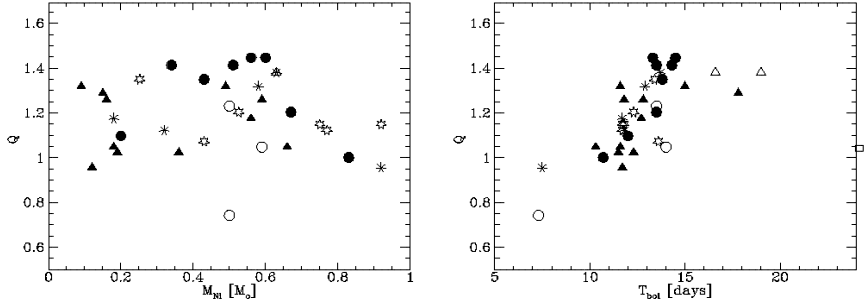,width=12.6cm,rwidth=4.4cm,angle=270}
\caption{
Ratio Q between the bolometric luminosity and the energy release by 
 gamma decay at $t_{bol} $
 as a function of the $^{56}Ni $ mass (left) and 
rise time to bolometric maximum $t_{bol}$. The different
symbols correspond to different explosion scenarios (see Table 1).}
\end{figure}
\begin{figure}
\psfig{figure=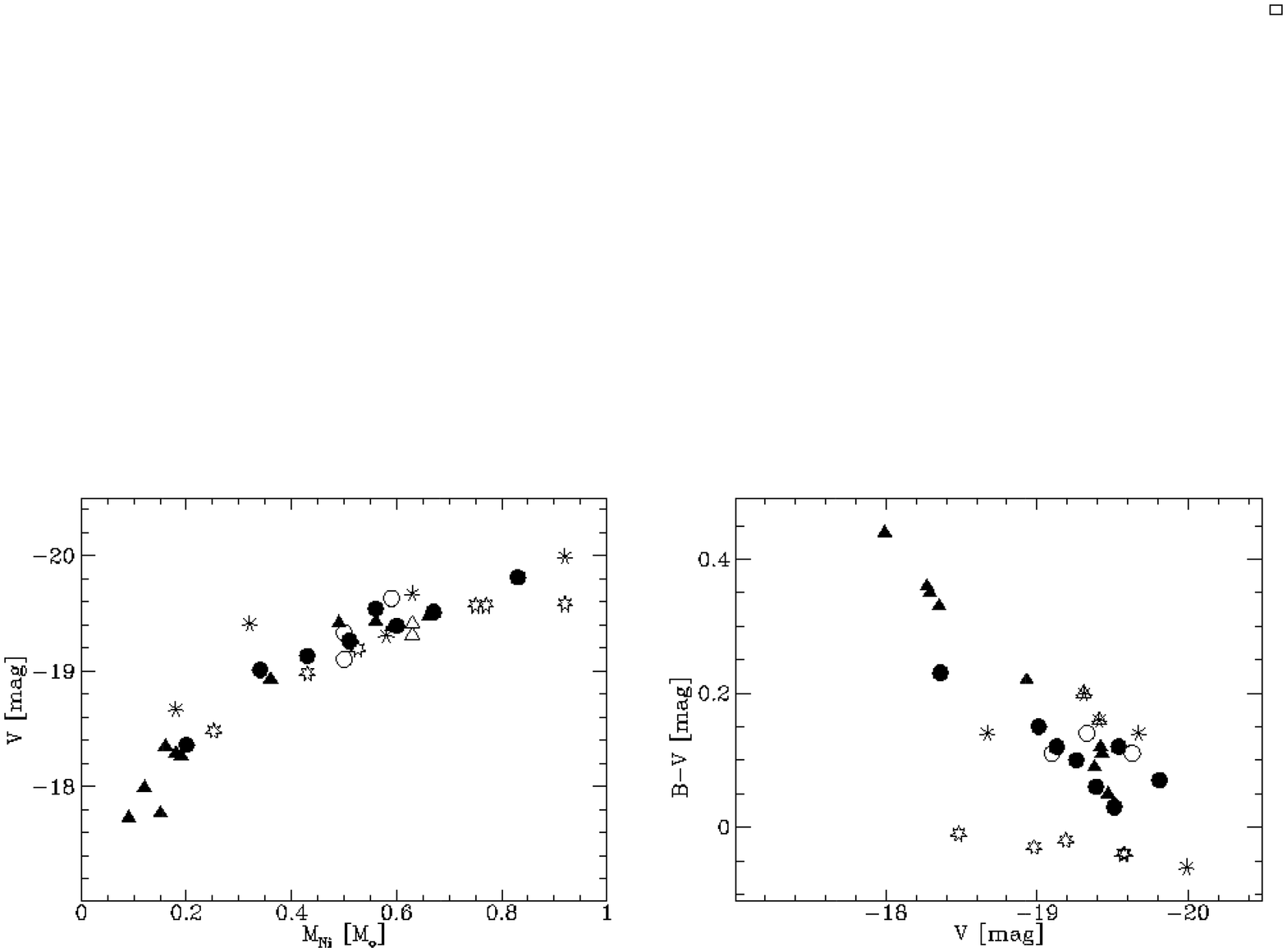,width=12.6cm,rwidth=4.4cm,angle=270}
\caption{
Absolute visual brightness V  as a 
function of the $^{56}Ni$ mass (left) and intrinsic color B-V 
 as a function of V(right).}
\end{figure}

\section{Comparison with Observations}
 The different explosion scenarios can generally be distinguished based on differences in
the slopes of the early
 monochromatic LCs (Fig. 6) and the expansion velocities indicated by the spectra
(Fig. 9 \& H\"oflich, this volume).
 For all models with a $^{56}Ni$ production  $\geq 0.4 
M_\odot $, $M_V$ ranges  from -19.1 to $-19.7^m$. 
 As a general tendency, the post-maximum declines
are related to $M_V$,  but there is a significant spread in the decline rate 
 among models with similar
brightess.  For all models but the Helium detonations, the colors become very red for  
small  $M_{Ni}$ (Fig. 8).
 
 As a general trend, the maxima of subluminous supernovae are more pronounced.
The reason becomes obvious if we consider the following related point.
In the literature (Arnett et al. 1985), it is often assumed that
the bolometric luminosity $L_{bol}$ is equal to the instantaneous release of energy by radioactive
decay $L_{\gamma}$, i.e. $ Q= L_{bol} / \dot E_{\gamma} \approx 1$, independent from the model.
 We find 
  that  $ Q$ depends on the model because the opacities vary strongly with  
temperature. Typically, $Q$ ranges from  0.7 to 1.4 (Fig. 7).
 For models with fast rising LCs, the opacity  stays high and 
the photosphere recedes mainly by the geometrical dilution of matter. For models with a slower
rise time or little $^{56}Ni$, the opacity drops strongly at about maximum light. The photosphere
receds quickly in mass, and thermal energy can be released from a larger region.
 Because no additional energy is gained, the energy reservoir is exhausted faster, and the
post-maximum  decline becomes steeper.
 
 The distinguishing features of Helium detonations are  blue 
 maxima  both for normal and subluminous
supernovae, a rapid increase of the luminosity
caused by $^{56}Ni$ heating, followed by a phase of slow rise to maximum light 
  and a 
fast  post-maximum decline. The fast decline is caused by the rapid increase of the
 escape probability of $\gamma  $ photons originating in the outer $^{56}Ni$ 
and the short life time of $^{56}Ni$. The early synthetic spectra are dominated by high velocity Ni and
the  absence of Si which is underabundant by more than  two orders of magnitude compared
to delayed detonation models.
 Note that a significant  amount of the line emission at late
phases should be powered  by $^{56}Co$ at  high velocities.
In  the subluminous models, about 1/2 to 1/4 of the radioactive material is ejected at high
velocities.  Accordingly, late-time spectra should show broad Co and Fe lines.
 
\section {Comparison between Observation and Models Predictions}
  
 We use  a quantitative method for fitting  data to models based on Wiener filtering 
(Rybicki and Press, 1995). 
The reconstruction technique is applied to the standard deviation from the theoretical LC to avoid
problems with measurements  distributed   unevenly in time.
By minimizing the error, 
the time of the explosion, the distance, and the reddening correction can be determined.
                                      For details, see H\"oflich \& Khokhlov (1995).
 
\begin{table}[htb]
\begin{center}
\caption{
 List of observed SNeIa used in our sample.
  Columns 2 to 7 give the parent
galaxy, its type,  peculiar velocity $v_z$ according to the MERCG catalogue (Kogoshvili 1986),
the distance modulus of the host galaxy and 
 the color excess $E_{B-V}$ of the SNe~Ia as determined from our models.
In column 8, we give
 names of those theoretical models (see Table~1) which can reproduce
the observed LCs. Models in bracketts do not fullfil our criterion but are
close. These models have been excluded for M-m. Models marked with $^1$ 
can be ruled out because the observations indicate Si at high velocities incompatible with the
chemical profiles. Note that the new delayed detonation models have not been included in this
list.
 We found some evidence for  a relation between the type of the explosion
and of the host galaxy. In elliptical galaxies, SN~Ia  with shell-like structures
are highly favored. These may be understood within the pulsating and merging scenarios.
 Because ellipticals consist on low mass stars only, this may provide a hint to the
progenitor evolution; however, a larger sample of observations is needed to confirm this trend.
}
\begin{tabular}{llllllll}
\hline 
 Supernovae  &  galaxy &Type~& D$ [Mpc]  $ & $E_{B-V}$& acceptable models                \\
\hline                     
 SN~1937C     & IC 4182  & Sm &  4.5$\pm 1 $   & 0.10 & N32,W7,DET2  \\ 
 SN~1970J     & NGC 7619 & dE & $ 63 \pm 8$     & 0.01 & DET2env2/4,(PDD3)  \\  
 SN~1971G     & NGC 4165 & Sb & $ 36 \pm 9 $    & 0.04  &  M37, M36, W7,N32 \\ 
 SN~1972E     & NGC 5253 & I  & 4.0$\pm 0.6 $   & 0.04 & M35,N21,M36,(HeD12$^2$)   \\ 
 SN~1972J     & NGC 7634 & SBO& $ 52 \pm 8 $    & 0.01 & W7,M36/37,N32,DET2   \\ 
 SN~1973N     & NGC 7495 & Sc & $ 69 \pm 20 $   & 0.08 & N32,M36,W7,(HeD10/11)   \\
 SN~1974G     & NGC 4414 & Sc & $ 17.5 \pm 5 $  & 0.0  & N32,M36,W7,(HeD10/11)  \\ 
 SN~1975N     & NGC 7723 & SBO& $28 \pm 7 $    & 0.18: &  PDD3/6/9/1a\\  
 SN~1981B     & NGC 4536 & Sb & 19 $\pm 4 $    & 0.05 &M35, N21  \\ 
 SN~1983G     & NGC 4753 & S  & $ 15 \pm 4 $    & 0.29 & N32, W7 (M36)\\ 
 SN~1984A     & NGC 4419 & Ep&  $ 16 \pm 4 $   & 0.14 & DET2env2, PDD3/6) \\ 
 SN~1986G     & NGC 5128 & I  &  4.2$\pm 1.2 $ & 0.83 & W7, N32, (M37/8)  \\ 
 SN~1988U     & AC 118$^1$ &  - &$1440 \pm 250 $   & 0.05 &     M36, W7, N32  \\ 
 SN~1989B     & NGC 3627 & Sb & $ 8.7 \pm 3 $   & 0.45 & M37, M36\\ 
 SN~1990N     & NGC 4639 & Sb & $ 20 \pm 5 $    & 0.05 & DET2env2/4, PDD3/1a \\
 SN~1990T     & PGC 63925 & Sa &$ 180 \pm  30 $    & 0.1 & M37, M38  \\ 
 SN~1990Y     & anonym.   & E & $  195\pm  45 $    & 0.05 & W7,N32,M36/37,PDD1c \\ 
 SN~1990af    & anonym.   & E &  $265\pm  85 $    & 0.05& W7, N32, M36 \\ 
 SN~1991M     & IC 1151  & Sb & $ 41 \pm 10 $   &  0.12    & M35,  PDD3\\
 SN~1991T     & NGC 4527 & Sb & $ 12 \pm 2 $    & 0.10 & PDD3/6/1a. DET2env2  \\ 
 SN~1991bg    & NGC 4374 &  dE &$ 18  \pm 5 $    &  0.25      & PDD5/1c \\
 SN~1992G     & NGC 3294 &  Sc &$ 29 \pm  6 $   &  0.05 & M36/35, PDD3, HeD10 \\ 
 SN~1992K     &ESO269-G57 & SBb &$43 ^{+15}_{-8} $   &  0.18 & PDD5/1a,(M39,HeD2)\\ 
 SN~1992bc    & ESO-G9   &  S  &$ 83 \pm 10 $   &  0.04 & PDD6/3/1c   \\ 
 SN~1992bo    & ESO-G57  &  S  &$ 79 \pm 10 $   &  0.03 & PDD8  \\ 
 SN~1994D     & NGC 4526 &  S0 &$ 16 \pm 2  $   &  0.00 & M36, (W7, N32)   \\
%
\hline 
\end{tabular}   
\end{center}
\end{table}

Observed monochromatic LCs and spectra of 27 SNe~Ia  are  compared with 
theoretical models (Table 2, Fig. 9 \&  H\"oflich, this volume).
According to our results, normal bright, fast SNeIa (e.g.  SN~1971G,
SN~1994D) with rise times up to 15~days (17~days) for the blue
(visual) LC can be explained by delayed detonation with
 different densities $\rho_{tr}$ for the transition from a deflagration to
a detonation. For PDDs, the density $\rho_{tr}$ stands for the density
at which the detonation starts after the
first pulsation. Typically, $\rho _{tr} $ is about $2.5~10^7~g~cm^{-3}$.
 Central densities of the initial WDs  range from 2.1 to 
   3.5 $10^9 g~cm^{-3}$. As a tendency, models at the lower end of this range tend
to give somewhat better fits. 
This may be explained by a high
accretion rate, some variation in the chemical composition or by an additional trigger 
mechanism for the explosion.
We note that the classical deflagration W7 (Nomoto et al. 1984)
provides similar good fits in several cases because its structure resembles
those of DD models but it has some problems with the high velocities of 
Si lines in SN1994D  (H\"oflich 1995).

\begin{figure}
\psfig{figure=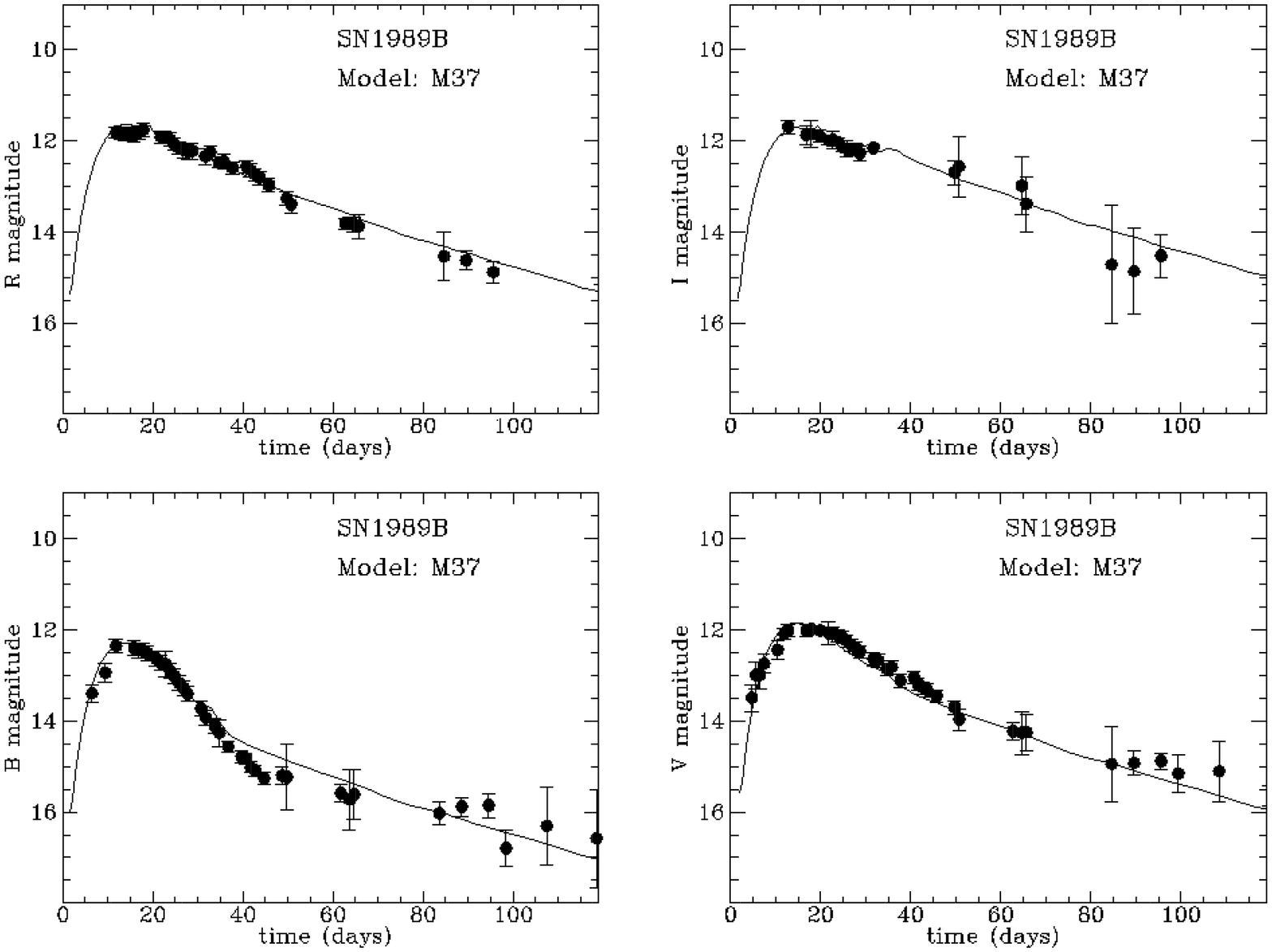,width=12.4cm,rwidth=4.4cm,clip=,angle=270}
\hskip 1.cm
\psfig{figure=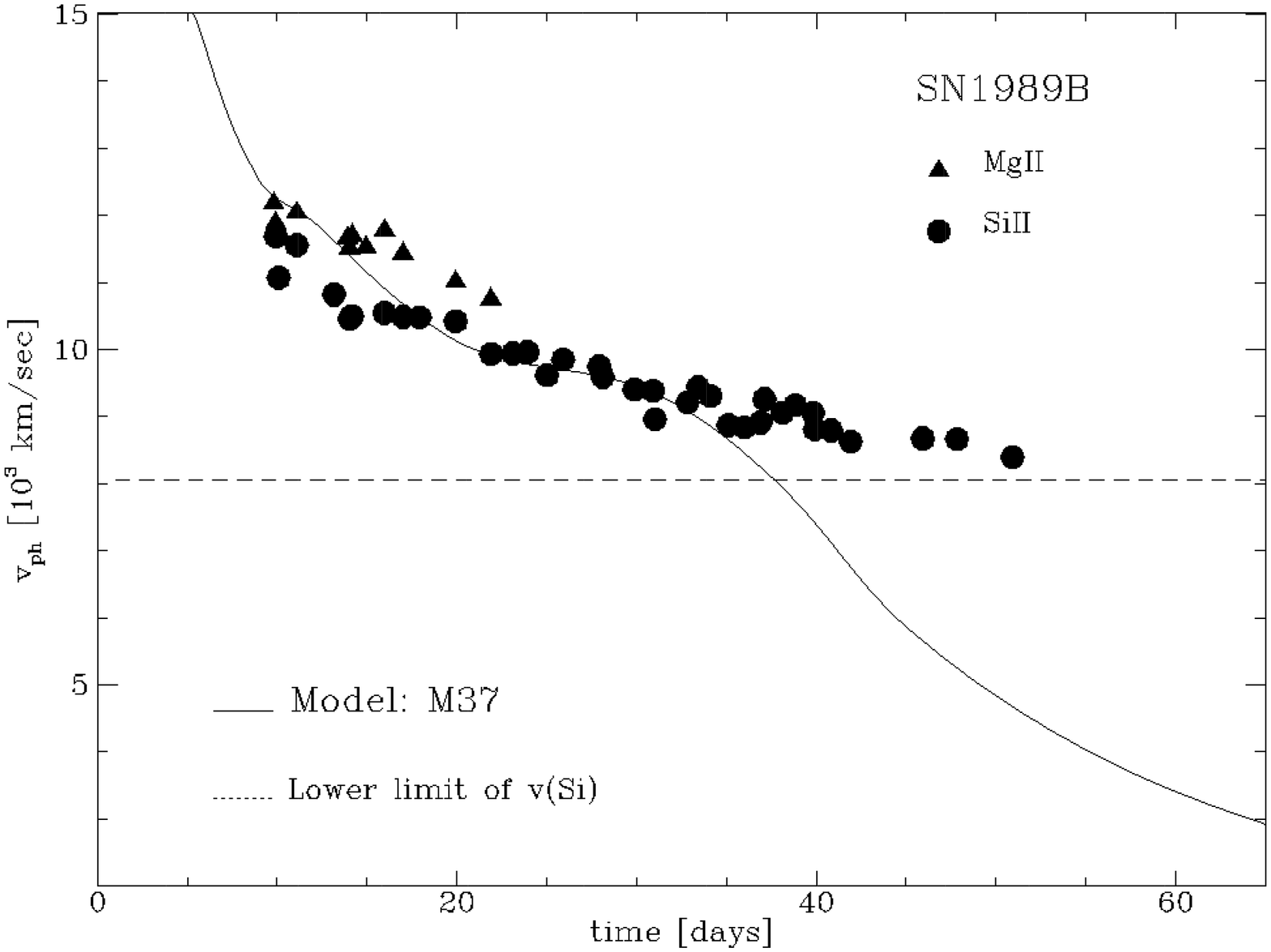,width=5.3cm,rwidth=4.4cm,clip=,angle=270}
\vskip -3.5cm
\hskip 6.1cm
\psfig{figure=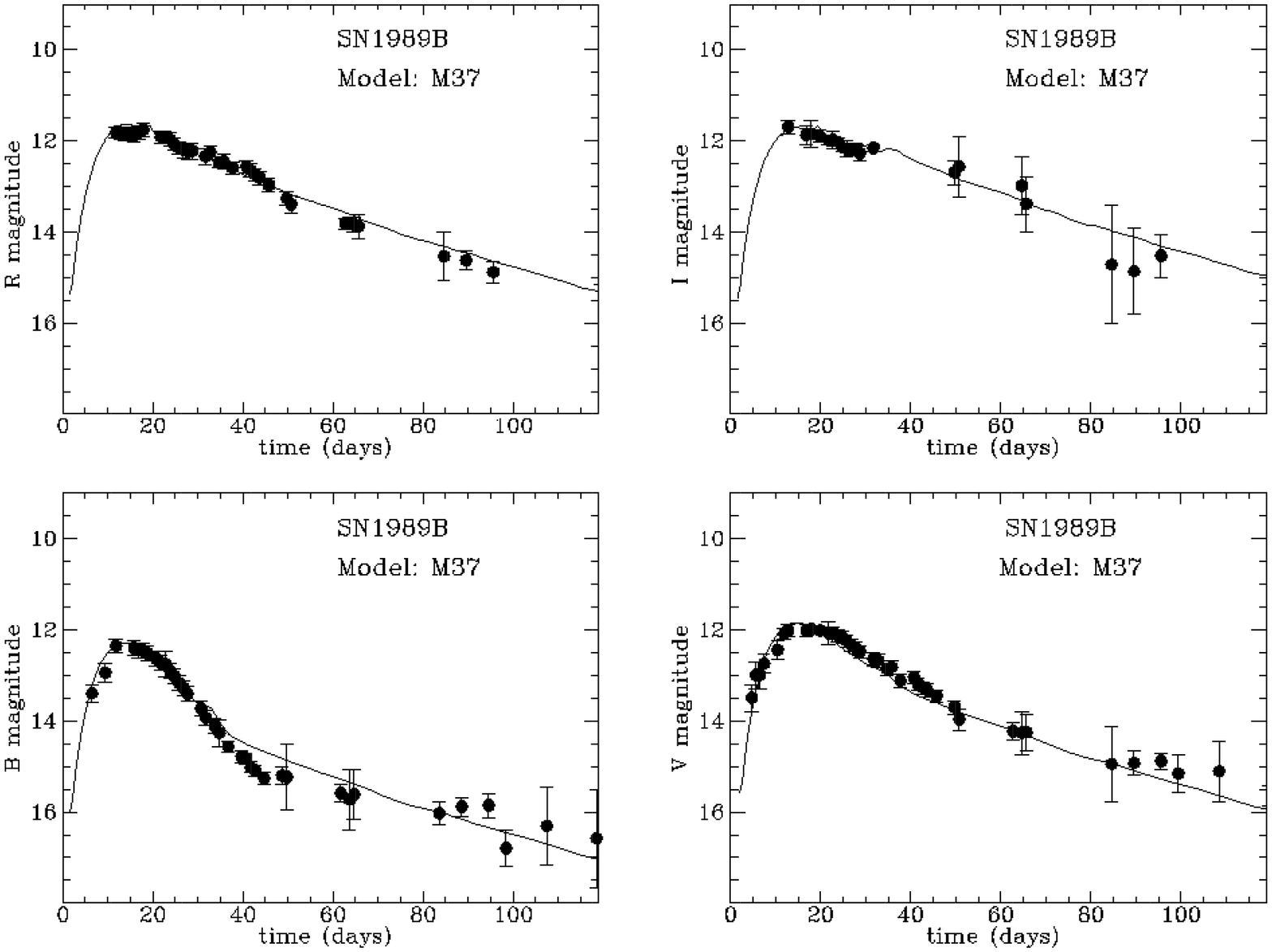,width=6.2cm,rwidth=4.4cm,clip=,angle=270}
\caption{
Monochromatic LCs of SN~1989B 
 band compared with the calculated LC of the
delayed detonation model M37. The 2 $\sigma $-error ranges are
given by Wells et al. (1994). In the lower left plot, the
 photospheric velocity of the delayed detonation model M37 is given in 
comparison with the expansion velocities implied by the line shifts of Mg II
and SiII. The dotted line is the velocity at which, in M37, the Si abundance has dropped
to 10 \% of its maximum value. The `leveling off' of the line shift is produced by
the chemical profile and not due to a change of the density profile. Note that detailed
spectral synthesis is required to archieve a better accuracy or to test the ionization
balance (H\"oflich 1995).
} 
\end{figure}

The ``standard" explosion models are unable, however, to reproduce rise
times to blue (visual) maximum longer than 15~days (17~days), provided
the progenitor is a C/O WD          of about 1.2 to $1.4 M_\odot$.     
  In fact, slow rising and declining LCs have been
observed (e.g.  SN~1990N, SN~1991T) which require models with an envelope of
typically 0.2 to $0.4M_\odot$. The envelope can be produced during a strong pulsation
or during the  merging of two WD.          
  The lower
value should not be regarded as a physical limit, because  it is likely
that a continuous transition from models with and without an
envelope exists. Note that a unique feature of models with massive envelopes are
very high photospheric expansion velocities ($v_{ph} \approx
16,000$~km/s) shortly before maximum light, which drop rapidly to an almost
constant value between 9000 and $12,000$~km/s.  This ``plateau" in
$v_{ph}$ lasts for 1 to 2~weeks depending on the envelope mass 
(Khokhlov et al. 1993).  In fact, there is some evidence for the plateau in
$v_{ph}$ from the Doppler shift of lines of SNe~Ia with       a slow
pre-maximum rise and post-maximum decline
              (e.g.  SN~1984A, SN~1990N,  M\"uller \& H\"oflich 1994).
 
Strongly subluminous supernovae (SN1991bg, SN1992K, SN1992bc) can 
be explained within the framework of pulsating
delayed detonation models with a low transition density. 
 In particular, the models become systematically redder and the post-maximum
decline becomes steeper with       decreasing 
brightness in agreement with observations. However, we do not get 
unique relations between these different quantities.
 The evolution
of the photospheric expansion velocity $v_{ph}$ (H\"oflich et al. 1995) and,
 in particular, its steady decline, is  consistent with observations.
 We must also note that we  need a rather high intrinsic reddening 
 for SN1992K and SN1991bg. Whether this can be explained by 
selective line blanketing, dust formation, or foreground clouds, or a combined effect 
 is under investigation.
 The latter possibility must be regarded as unlikely, because, at later phases, the colors
of these two supernovae are close to those of bright SNe~Ia.
Note that some of the `classical' delayed detonation models (M39) also produce
strongly subluminous LCs, but these do not fit  any of the measurements.     
 
Our Helium detonation models are rather unsuccessful in reproducing observations,
mainly due to the rather steep post-maximum  declines  for normal bright supernovae, and
Si lines observed at higher expansion velocities than compatible with the chemical 
profiles. At early times,
 our synthetic spectra are dominated by Fe-group elements and lack strong Si features 
because of the low abundances of Si ($\leq 10^{-2...-3}$).
 For subluminous supernovae,
the blue color at maximum light and the strong IR-maximum are both in contradiction to the
observations. 
 Quantitatively, multi-dimensional effects may alter the LCs mainly due to an increased   
escape probability for photons.
  However, the basic features of the 
LCs and spectra are  not expected to change 
because they are inherent to the outer $^{56}Ni$. The  energy required to push
the entire $^{56} Ni$ to sufficient high velocities ($\leq 16,000 km/sec$)
 is well in excess of the entire energy production.
  
\begin{figure}
\psfig{figure=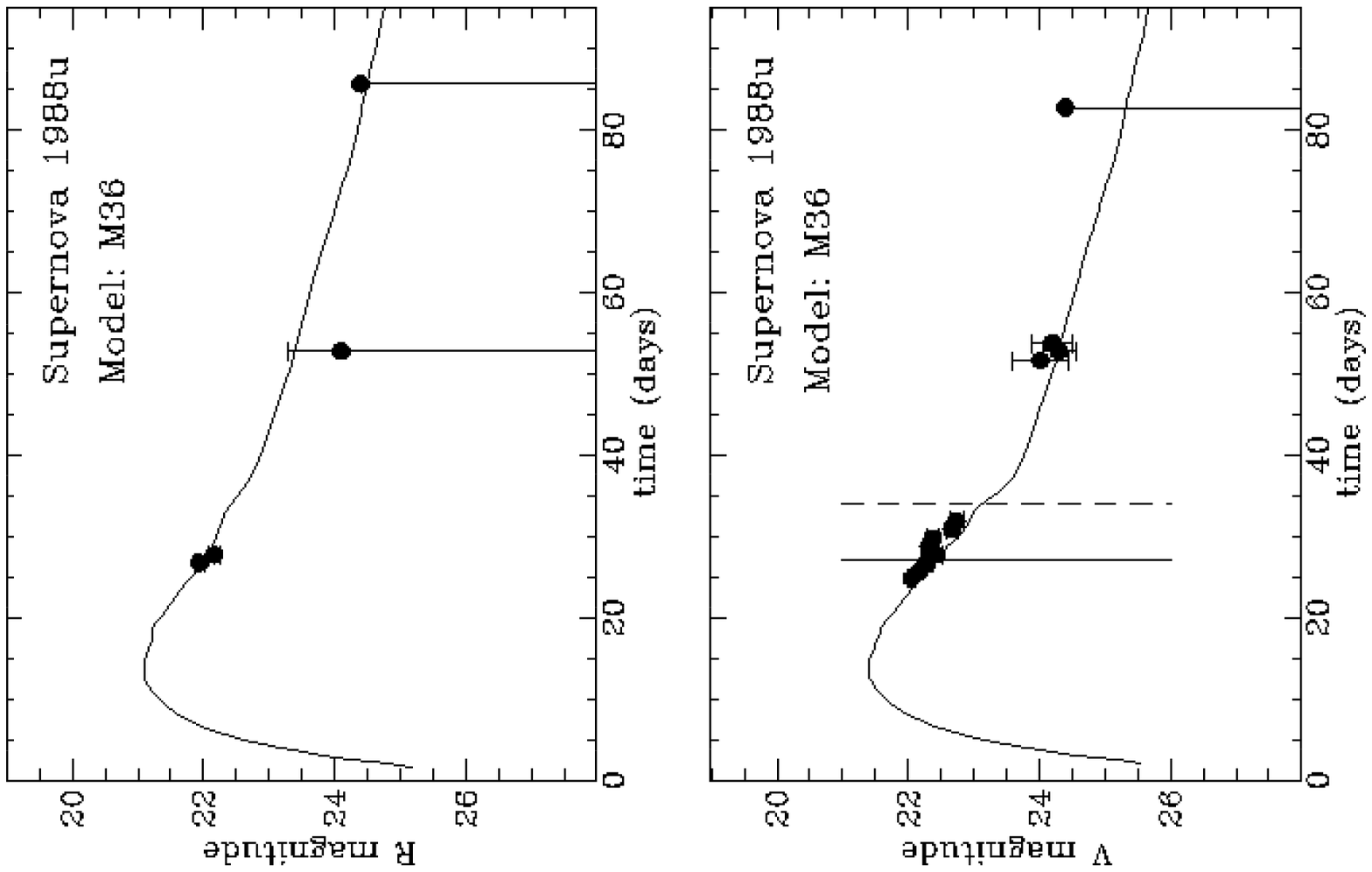,width=12.4cm,rwidth=4.4cm,clip=,angle=270}
\vskip -4.0cm
\psfig{figure=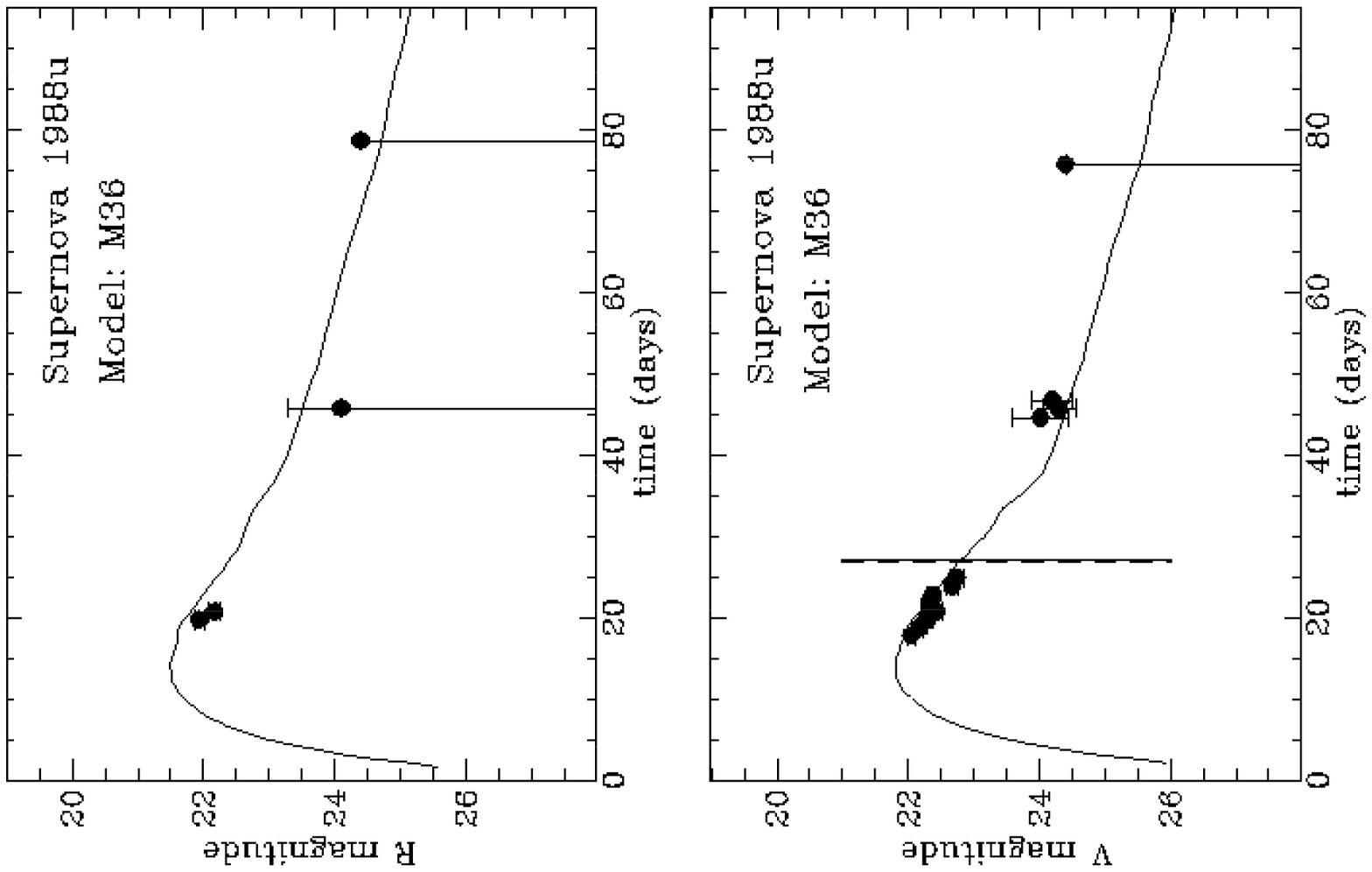,width=6.2cm,rwidth=4.4cm,clip=,angle=270}
\caption{
 Visual LC  of SN~1988U           
in comparison with the calculated LC of the
delayed detonation model M36. Both the V and R colors are used.
The dashed line marks the time
when the photospheric velocity of the models corresponds to the observed line shift
(thin line). The right and left plots correspond to different time shifts and
both allow for a reasonable reproduction of the LC. This would correspond
to an uncertainty  of $0.5^m$ in the brightness. However, 
the right plot can be ruled out by       the spectrum. This clearly demonstrates
the importance of simultaneous analysis of both LCs and spectra.  
 }
\end{figure}
 
 Our findings with respect to the explosion scenario can be concluded as follows.
 Models with masses close to the Chandrasekhar limit provide
the best agreement with the observations.
 Delayed-detonation and deflagration models similar
to W7 and pulsation or merging scenarios are required. Based on the sparse observations,
we cannot exclude the mechanism of Helium detonations. In particular, more early time
spectra are crucial.           
 
 Pinto's group end up with a different result.  Based on their 
calculations of the delayed detonation model DD4, they exclude this scenario. Their 
 B and V LCs show rise times in excess of 21 days whereas both the U and 
bolometric LCs peak around day 15 and, at maximum light,
their optical spectra  show Co 
lines in strong emission (talk by R. Eastman). However, according to Pinto
(priv. communication), 
 their Helium detonations provide both
excellent fits to spectra and LCs
without encountering our Si problem.
%
%
 We find $H_o $ to be $67 \pm 9 km/sec Mpc$ within a 95 \% confidence level.
This value agrees well with our previous analysis based on a subset of observations
and models ($66 \pm 10 km~Mpc^{-1} sec^{-1}$, M\"uller \& H\"oflich 1994).
 A strong variation of the local value can be 
ruled out at least on  scales below 200 Mpc. From SN1988U, the deceleration
parameter $q_o$ is $0.7 \pm 1.0 $ ($2  \sigma $). Better limits can we expected in the near
future when more and
 more distant SNeIa become available
(Perlmutter et al. 1995).

\section{Distance determinations, $H_o$ and $q_o$}
 
Based on our LCs, we have also determined the individual
distances of the parent galaxies of the analyzed SNe~Ia (Table 2). Our method           
does not rely on secondary distance indicators and allows for a consistent
treatment of interstellar reddening and the interstellar redshift.  The
advantages of a consistent inclusion of information from the spectra becomes striking
for SN1988U (Fig. 10).
\begin{figure}
\psfig{figure=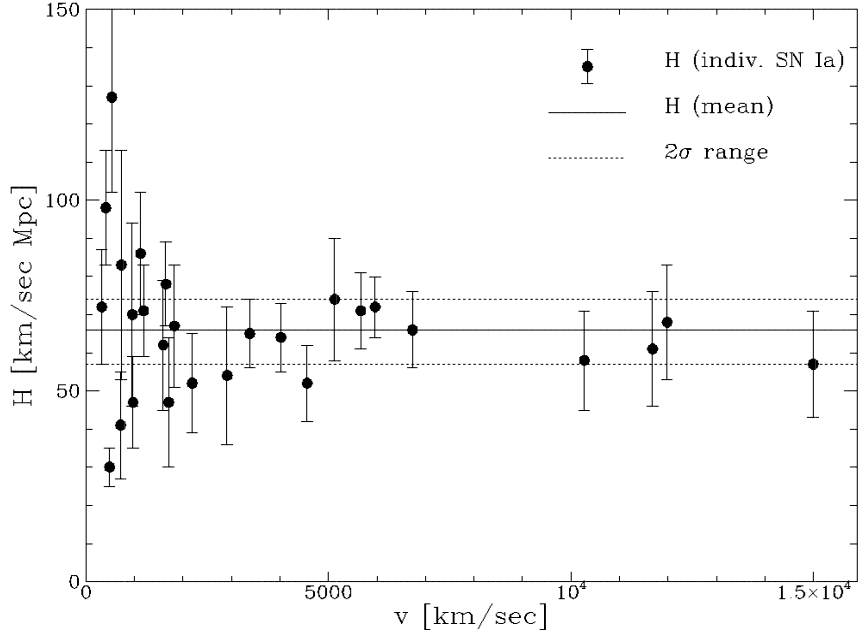,width=12.6cm,rwidth=6.5cm,angle=270}
\caption{
Hubble values H are shown based on individual distances (see table 2).
 SN1988U at v=91500 km/sec gives $H= 64 \pm 10$  [km/sec Mpc].
}
\end{figure}
  
 Other determinations of $H_o$ which are based on independent,
purely statistical methods and primary distance indicators. It may be encouraging that the result 
of different SN~Ia based methods agree
 if SN~Ia are not treated as as  standard candles.
    Hamuy et al. (1995)  found
$65 \pm 5$,  Riess et al. (1995) give $67 \pm 5$,  and 
Fisher et al. (1995), Nugent et al. (1995b) and Branch et al, (1995)
 get a values of $60 \pm 10$, $60 \pm 12$ and 
$58 \pm 7$, respectively.
From our models, both the empirical relations between
 $M_V/dM(15)$ like-relations and the ansatz to deselect subluminous SNeIa seems 
to be justified, but we expect an individual  dispersion  of $\approx 20 \%$ 
(Hamuy et al. 1995b).

%

%

\parindent=0pt
\baselineskip=16pt
\section{References}
\bibliographystyle{}
\begin{small}

Arnett, W.D., Branch D., Wheeler J.C. (1985), {\it Nature}, {\bf 314}, 337
\hfill

Arnett, W.D. (1969), {\it ApJS},{\bf 5}, 180
\hfill

Athay R., (1972), {Radiation Transport in Spectral Lines},
{Dordrecht}, {D.Reidel Publ.Comp.}
\hfill

Ayres T., Wiedemann G.R. (1989), {\it ApJ}, {\bf 338}, 1033
 \hfill

 {Barbon R., Bennetti S., CApJalaro E., Rosino L., Turatto M.}  (1990),
 {\it A\&A}, {\bf 237}, {79}
\hfill

 {Bessel, M.}, ({1990}),{\it PASP}, {\bf 102}, {1181}
\hfill

 {Branch  D., et al.}  (1995), {\it Phys. Rep.}, submitted
\hfill

{Canal R.}  ({1995}),{Les Houches Lectures},
eds. J. Audouze et al., Elsevier, in press
\hfill
 
Castor, J.I. (1974), {\it MNRAS}, {\bf 169}, 279
\hfill

{Collela  P., Woodward  P.R.} (1984),{\it J.~Comp.~Phys.}, {\bf 54}, 174
\hfill

{Filippenko  A.V. et al.}
         (1992), {\it AJ}, {\bf 104}, 1543
\hfill

 {Fisher  A., Branch  D., H\"oflich P., Khokhlov A., Wheeler J.C}, (1995), {\it ApJ}, {\bf 447L},  731
\hfill

{Frogel J.A. et al.}
(1987), {\it  ApJ}, {\bf 315L}, 129
\hfill

 {Hachisu I., Eriguchi Y., Nomoto K.} ({1986b}), {\it A\&A}, {\bf 168}, {130}
\hfill

 {Hamuy M. et al.} ({1994}), {\it AJ}, {\bf 108}, {2226}
\hfill

 {Hamuy M. et al.} ({1995}), {\it AJ}, {\bf 109}, {1}
\hfill
 
  {Hamuy M., H\"oflich P., Khokhlov A., Wheeler J.C.}  (1995b), {\it ApJ}, accepted
\hfill

Hansen, C.J., Wheeler, J.C. (1969) {\it ApJSS}, {\bf 3}, 464
\hfill

Harkness, R.P. (1991), in: SN1987A, ed. I.J. Danziger, ESO, Garching, p.447
\hfill
 
H\"oflich, P., (1990), Habilitation thesis, U. of Munich, MPA 563
\hfill

H\"oflich, P., Khokhlov, A., M\"uller, E. (1991),{\it  A\&A},{\bf 248}, L7
\hfill

 {H\"oflich, P., M\"uller, E,; Khokhlov, A.} (1993), {\it A\&A}, {\bf 268}, {570} 
\hfill

 {H\"oflich  P., Khokhlov  A. }  (1995), {\bf ApJ}, {in press}
\hfill

 {H\"oflich  P., Khokhlov  A., Wheeler  C. }  (1995), {\it ApJ}, {\bf 444}, 831
\hfill

 {Hoyle  P., Fowler }, ({1960}) {\it ApJ}, {\bf 132}. 565
\hfill

Iben, I.Jr., Tutukov, A.V. (1984), {\it ApJS}, {\bf 54}, 335 
\hfill

Jeffery D. (1995), {\it A\& A}, {\bf 299}, 770
\hfill

 {Khokhlov  A., M\"uller  E., H\"oflich  P.} (1993), {\it A\&A}, {\bf  270}, {23}
\hfill

 {Khokhlov  A.}, (1991ab), {\it A\&A}, {\bf 245}, {114} \& {L25}
\hfill

 {Leibundgut, B. et al.}, ({1993}), {\it AJ}, {\bf 105}, {301}
\hfill

Karp, A.H., Lasher, G., Chan, K.L., Salpeter E.E., 1977, {\it ApJ }, {\bf 214}, 161
\hfill

    {Kurucz, R.} (1993),{Atomic Data for Opacity Calculations,} {Cambridge,}{ CfA}
\hfill

Livne  E., Arnett D. (1995), {\it ApJ}, in press
\hfill

Mihalas, D. (1978), Stellar Atmospheres, Freeman, San Francisco
\hfill

 {Mihalas D., Kunasz R.B., Hummer D.G.} ({1975}),   {\it ApJ}, {\bf 202}, 465
\hfill

Mihalas D., Kunasz R.B., Hummer D.G. (1976),  {\it ApJ}, {\bf 206}, 515
\hfill

Mihalas D., Kunasz R.B., Hummer D.G. (1976b),  {\it ApJ}, {\bf 210}, 419
\hfill

 {M\"uller  E., H\"oflich  P.}, ({1994}), {\it A\&A}, {\bf 281}, {51}
\hfill

{ Nomoto  K., Yamaoka  H., Shigeyama  T., Iwamoto  K.}, ({1995}), 
{\sl in: Supernovae}, ed. R.A. McCray, {Cambridge University Press}, in press
\hfill

 {Nomoto K., Thielemann F.-K., Yokoi K.} ({1984}), {\it ApJ}, {\bf 286}, 644
\hfill

 {Nomoto  K., Sugimoto  D.}, ({1977}), {\it PASJ}, {\bf 29},{ 765}
\hfill

 { Nomoto  K.} ({1982}), {\it ApJ}, {\bf 253}, 798
\hfill

Nomoto, K., Sugimoto, D., Neo, S. (1976) {\it  ApJSS}, {\bf 39}, 137
\hfill

 {Nugent P., Baron E.,Hauschild P., Branch D.} (1995a), {\it ApJ}, {\bf 441L}, 33
\hfill

 {Nugent P., et al.} (1995b), 
{\it Phys. Rev. Let.}, {\bf 75}, {394 \& 1974(E)}
\hfill

  {Olsson G.L., Auer L.H., Buchler J.R.} (1986), {\it JQSRT}, {\bf 35}, {431}
\hfill

  {Perlmutter et al. } (1995), {\it ApJ}, {\bf 440L}, 41
\hfill

{Phillips  M.M. et al.}(1987),  {\it PASP}, {\bf 90}, {592}
\hfill

 {Pskovskii  Yu.P.}, {(1970)},{ \it Astron. Zh.}, {\bf 47}, {994}
\hfill

 {Riess A.G., Press W.H., Kirshner R.P.}, {(1995)},{ \it ApJ}, {\bf 438L}, 17
\hfill

{Ruiz-Lapuente  P.  et al.   }, ({1993}), {\it Nature}, {\bf 365}, {728}
\hfill

 {Rybicki G.B., Press W.H.} (1995), {\it Phys.Rev.Let.}, {in press}
\hfill

 {Shigeyama T. et al} (1993), {\it AAS}, {\bf 97}, 223  
\hfill

{Thielemann  F.-K. et al.} ({1994}),
{in: Supernovae}, {Elsevier}, {Amsterdam}, {629}
\hfill

{ Thielemann  F.-K., Arnould  M., Truran  J.W.}, ({1987}),
{ Advances in Nuclear Astrophysics}, {E. Vangioni-Flam}{Editions fronti\`eres}, {Gif sur Yvette}, {525}
\hfill

Wells, L.A. et al. (1994), {\it AJ}, {\bf 108}, 2233
\hfill

{Wheeler  J. C., Harkness  R .P.}, ({1990}), {\it Rep. Prog. Phys.}, {\bf 53}, {1467}
\hfill

{Wheeler J. C., Harkness R .P., Khokhlov A., H\"oflich P.}, (1995),
{\it Phys.Rep.}, in press
\hfill

{Woosley  S. E., Weaver T. A.}  ({1994}), {in: Supernovae}, {Elsevier}, {Amsterdam}, {423} 
\hfill

 {Woosley  S. E. \& Weaver, T. A.} (1994), {\it ApJ}, {\bf 423}, 371 
\hfill

 {Yamaoka H., Nomoto K., Shigeyama T., Thielemann F.} {(1992)}, {\it A\&A}, {\bf 393}, 55
\hfill
 
\end{small}

\end{document}